\documentclass[sn-nature, pdflatex]{sn-jnl}

\usepackage{graphicx}%
\usepackage{multirow}%
\usepackage{amsmath,amssymb,amsfonts}%
\usepackage{amsthm}%
\usepackage{mathrsfs}%
\usepackage[title]{appendix}%
\usepackage{xcolor}%
\usepackage{textcomp}%
\usepackage{manyfoot}%
\usepackage{booktabs}%
\usepackage{algorithm}%
\usepackage{algorithmicx}%
\usepackage{algpseudocode}%
\usepackage{listings}%
\theoremstyle{thmstyleone}%
\theoremstyle{thmstyletwo}%

\theoremstyle{thmstylethree}%

\begin{document}

\title[Redox and Electronic structure]{Direct Evidence of Metal-Ligand Redox in Li-ion Battery Positive Electrodes}

%%=============================================================%%
%% GivenName	-> \fnm{Joergen W.}
%% Particle	-> \spfx{van der} -> surname prefix
%% FamilyName	-> \sur{Ploeg}
%% Suffix	-> \sfx{IV}
%% \author*[1,2]{\fnm{Joergen W.} \spfx{van der} \sur{Ploeg} 
%%  \sfx{IV}}\email{iauthor@gmail.com}
%%=============================================================%%

\author*[1,2]{\fnm{Galo J.} \sur{P\'aez Fajardo}}\email{galo.paez-fajardo@warwick.ac.uk}
\equalcont{These authors contributed equally to this work.}

\author[1]{\fnm{Daniela} \sur{Dogaru}}%\email{iiauthor@gmail.com}
\equalcont{These authors contributed equally to this work.}

\author[3,4]{\fnm{Hrishit} \sur{Banerjee}}%\email{iiauthor@gmail.com}

\author[1,2]{\fnm{Muhammad} \sur{Ans}}
\author[1,2]{\fnm{Matthew J. W.} \sur{Ogley}}
\author[1,2]{\fnm{Veronika} \sur{Majherova}}
%\author[1]{\fnm{Gaurav C.} \sur{Pandey}}
%\author[1,2]{\fnm{Ashok S.} \sur{Menon}}
\author[2,5]{\fnm{Innes} \sur{McClelland}}
\author[6]{\fnm{Shohei} \sur{Hayashida}}
\author[6]{\fnm{Pascal} \sur{Puphal}}
\author[6]{\fnm{Masahiko} \sur{Isobe}}
\author[6]{\fnm{Bernhard} \sur{Keimer}}
\author[7]{\fnm{Pardeep K.} \sur{Thakur}}
\author[7]{\fnm{Tien-Lin} \sur{Lee}}
\author[7]{\fnm{Dave C.} \sur{Grinter}}
\author[7]{\fnm{Pilar} \sur{Ferrer}}
\author[2,5]{\fnm{Serena A.} \sur{Cussen}}
\author[6]{\fnm{Matthias} \sur{Hepting}}
\author*[1,2]{\fnm{Louis F. J.} \sur{Piper}}\email{louis.piper@warwick.ac.uk}

\affil*[1]{\orgdiv{WMG}, \orgname{University of Warwick}, \orgaddress{%%\street{Street},
 \city{Coventry}, \postcode{CV4 7AL}, %%\state{State}, 
\country{United Kingdom}}}

\affil[2]{%%\orgdiv{Department}, 
\orgname{The Faraday Institution}, \orgaddress{\street{Quad One, Harwell Science and Innovation Campus}, \city{Didcot}, \postcode{OX11 0RA}, %%\state{State},
 \country{United Kingdom}}}
 
\affil[3]{\orgdiv{School of Science and Engineering}, \orgname{University of Dundee}, \orgaddress{\street{Nethergate},
 \city{Dundee}, \postcode{DD1 4HN}, %%\state{State}, 
\country{United Kingdom}}}

\affil[4]{\orgdiv{School of Metallurgy and Materials}, \orgname{University of Birmingham}, \orgaddress{\street{Edgbaston},
 \city{Birmingham}, \postcode{B15 2TT}, %%\state{State}, 
\country{United Kingdom}}}

\affil[5]{%%\orgdiv{Department}, 
\orgname{University College Dublin}, \orgaddress{\street{Belfield},
 \city{Dublin}, \postcode{4}, %%\state{State},
 \country{Ireland}}}

\affil[6]{%%\orgdiv{Department}, 
\orgname{Max Planck Institute for Solid State Research}, \orgaddress{\street{Heisenbergstra{\ss}e 1}, \city{Stuttgart}, \postcode{70569}, \state{Baden-W\"urttemberg},
 \country{Germany}}}
 
\affil[7]{\orgdiv{Diamond Light Source Ltd.}, \orgname{Harwell Science and Innovation Campus}, \orgaddress{%\street{Torrington Place},
 \city{Didcot}, \postcode{OX11 0DE}, %%\state{State}, 
\country{United Kingdom}}}

%\affil*[1]{\orgdiv{Department}, \orgname{Organization}, \orgaddress{\street{Street}, \city{City}, \postcode{100190}, \state{State}, \country{Country}}}

%\affil[2]{\orgdiv{Department}, \orgname{Organization}, \orgaddress{\street{Street}, \city{City}, \postcode{10587}, \state{State}, \country{Country}}}

%\affil[3]{\orgdiv{Department}, \orgname{Organization}, \orgaddress{\street{Street}, \city{City}, \postcode{610101}, \state{State}, \country{Country}}}

%%==================================%%
%% Sample for unstructured abstract %%
%%==================================%%
\abstract{Describing Li-ion battery positive electrodes in terms of distinct transition metal or oxygen redox regimes can lead to confusion in understanding metal-ligand hybridisation, oxygen dimerisation, and degradation. There is a pressing need to study the electronic structure of these materials and determine the role each cation and anion plays in charge compensation. Here, we employ transition metal L-edge X-ray Resonance Photoemission Spectroscopy in conjunction with Single Impurity Anderson models, Self-consistent Real Space Multiple Scattering spectral simulations, and Dynamical Mean-Field theory calculations to directly evaluate the redox mechanisms in (de-)lithiated battery electrodes. This approach reconciles the redox description of two canonical positive electrodes---LiMn$_{0.6}$Fe$_{0.4}$PO$_{4}$ and LiNiO$_{2}$---in terms of varying degrees of charge transfer using the established Zaanen-Sawatzky-Allen framework, common to condensed matter physics. In LiMn$_{0.6}$Fe$_{0.4}$PO$_{4}$, the absence of charge transfer means capacity arises due to the depopulation of metal \textit{3d} states, i.e. conventional metal redox. Whereas, in LiNiO$_{2}$, charge transfer dominates and redox occurs through the formation and elimination of ligand hole states. This work clarifies the role of oxygen in Ni-rich system and provides a framework to explain how capacity can be extracted from oxygen-dominated states in highly covalent systems without needing to invoke dimerisation.}
%P 

\keywords{Li-ion battery positive electrodes, Electronic structure, X-ray spectroscopy, X-ray simulations}

%%\pacs[JEL Classification]{D8, H51}

%%\pacs[MSC Classification]{35A01, 65L10, 65L12, 65L20, 65L70}

\maketitle

%\section{Introduction}\label{sec1}

Oxygen loss limits the performance of both Li-excess and stoichiometric Ni-rich positive electrodes. \cite{Frith2023,Paez_2023,Zhang2022} Efforts to suppress it are often conflated with oxygen redox, which is considered to involve O$_2$ formation in Li-excess compounds. \cite{Marie2024} This has prompted the field to re-examine oxygen participation in redox more broadly, leading to two conceptually distinct formalisms over the last decade: non-bonding O \textit{2p} orbitals and metal-ligand orbital hybridisation. \cite{Menon_OxygeReview} The former has been widely used to explain O$_2$ formation observed in RIXS experiments, \cite{Marie2024} yet compounds that challenge this interpretation also show oxygen-dimer features at the top of charge. \cite{Hu2021, Zach_LP-topOfChargeC9MH00765B, Menon_2023PRXEnergy.2.013005, LNO_TrappedD3EE04354A} Recent studies further question whether oxygen-dimer RIXS signals correlate with reversible capacity \cite{Gao2025_Tarsacon_LPsug} while others propose alternative structural pathways to the O$_2$ formation as the dominant factor of the electrochemistry of Li-excess layered materials. \cite{Qiu2025_Shirley-LPsug, GENT20201369} Within this conflicting view, oxygen participation via hybridisation is increasingly supported, with beyond density function theory studies confirming its role to deliver capacity from the onset of delithiation, \cite{GenreithSchriever2023} validated by in operando X-ray measurements \cite{Ogley2025}. As a result, directly probing oxygen participation through orbital rehybridisation can thus inform computational models and reconcile conventional redox frameworks with this emerging hybridisation perspective, advancing our understanding of redox mechanisms.

\begin{equation} Li^+Ni^{3+}O_2^{2-}\rightarrow Li^++e^-+Ni^{4+}O_2^{2-}\label{eq:Nioxidation} 
\end{equation} 

Equation \ref{eq:Nioxidation} illustrates the end points of LiNiO$_{2}$ delithiation, where charge compensation raises the Ni oxidation state. Linking this evolution to the underlying electronic structure is essential to understand oxygen loss. Generally, the classical redox model ties oxidation states to changes in transition metal (TM) \textit{3d} charge density, assuming a fully ionic character of the TM-O bonds. \cite{Walsh2018, Pavarini:819465} This framework views oxidation as discrete changes in TM \textit{3d} orbital occupancy. \cite{book_ch3} LiMn$_{0.6}$Fe$_{0.4}$PO$_4$ is a canonical example of such ionic redox. 

\begin{equation} Li^{+}Fe^{2+}_{0.4}Mn^{2+}_{0.6} \left(PO_{4}\right)^{3-}\rightarrow Li^{+}+e^-+Fe^{3+}_{0.4}Mn^{3+}_{0.6}\left(PO_{4}\right)^{3-}\label{eq:Fexidation} 
\end{equation} 

Charge compensation in Equation \ref{eq:Fexidation} produces a Fe$^{2+}$/Fe$^{3+}$ redox pair that directly reflects variations in Fe \textit{3d} occupancy. By contrast, growing experimental evidence shows that this ionic picture fails to describe LiNiO$_{2}$ because of its more complex electronic structure shaped by bond-disproportionation, \cite{PhysRevB.100.165104} intersite charge fluctuations, \cite{Experiment_Holes_ESRF} and strong metal-ligand hybridisation. \cite{PhysRevB.109.035139} Under these influences, LiNiO$_{2}$ redox pathways deviate from the simple ionic view. Theoretical work has begun to address such complexities \cite{Melot_2013, GENT20201369, Assat2018}, but further experimental validation is needed to establish a general framework that captures electronic intricacies across cathode materials and reconciles emerging and conventional redox views, with particular attention to oxygen's role in charge compensation.

Most studies on Ni-rich layered compounds to date infer the redox mechanism indirectly, i.e. by probing core-level or unoccupied orbitals that are not themselves directly involve in charge compensation. In the classical ionic redox view, this indirect approach is generally sufficient. However, in compounds such as LiNiO$_{2}$, where the electronic structure is more complex, such indirect interpretations may fail, giving room to misleading conclusions. For instance, during Li deintercalation of LiNiO$_{2}$ positive electrodes, hard X-ray photoelectron spectroscopy (HAXPES) shows that Ni \textit{2p} peaks narrow but do not shift in energy as would be expected for ionic compounds. \cite{depth_resolvingACSenergy} In Ni K-edge X-ray absorption near-edge spectroscopy (XANES), the shift to higher energies of the raising edge---typically associated with Ni oxidation---halts despite continued capacity extraction, inconsistent with a model based solely on Ni oxidation. \cite{D4EE02398F, Ans2025, Menon_2023PRXEnergy.2.013005} Raman spectroscopy likewise reveals evolving E${g}$ and A$_{1g}$ peak intensities that suggest a continuous change in Ni-O bond covalency rather than the spectral behaviour characteristic of ionic bonds. \cite{Experiment_Holes_ESRF} Finally, Ni L-edge resonant inelastic X-ray scattering (RIXS) indicates a dominant Ni \textit{3d}$^{8}$ character, rather than the expected \textit{3d}$^{7}$ configuration implied by classical ionic redox. \cite{Experiment_Holes_ESRF} Taken together, these results highlight the need for methods that circumvent the limitations of indirect electronic probes and instead directly resolve the character and evolution of redox-active orbitals. Such an approach could advance our understanding of redox mechanisms, consistent with the emphasis in recent computational studies. \cite{Banerjee2024} Motivated by this need, our work implements a methodology that directly interrogates the orbitals involved in charge compensation. By unambiguously identifying their elemental and orbital character, this method avoids the concerns of indirect probes and provides a clearer understanding of the redox mechanism in electronically complex materials.

This approach is TM L-edge X-ray Absorption Spectroscopy (XAS)-assisted Resonant Photoemission Spectroscopy (RPES). \cite{Davis_PhysRevB.25.2912, PhysRevB.53.10372} By performing valence-band photoemission spectroscopy with X-ray energies tuned to TM L-edge transitions, RPES directly reveals which elements participate in redox and---when combined with the correct electronic assignment---how their electronic structure evolves. This assignment is provided by Single Impurity Anderson (SIA) models \cite{Haverkort_MLFT_2012}, which determine whether the selected transitions involve simple \textit{t$_{2g}$}/\textit{e$_{g}$} states or more complex features such as intersite charge fluctuations, bond-disproportionation, or metal-ligand hybridisation, thereby ensuring that resonant enhancement amplifies the relevant orbital character in the spectra. This reciprocal experiment-theory synergy enables direct probing of how intricate electronic structures dictate the redox mechanism, providing a material-agnostic framework for tracking orbital-level redox processes across diverse compounds and states of charge (SoC).

Here, we used the TM L-edge XAS-assisted RPES approach to directly study charge compensation mechanisms in the context of the simple ionic picture and more complex electronic structures. We used this approach on two canonical systems, LiMn$_{0.6}$Fe$_{0.4}$PO$_{4}$ (LMFP64) and LiNiO$_{2}$ (LNO). The high quality and electrochemical performance of the LMFP64 material and electrodes used in this work have been proven previously. \cite{Gerard_energyfuels} The LNO electrodes of this work exhibit the expected accelerated electrochemical performance decay \cite{Ans2025} as a consequence of cycling above its O-loss cell potential threshold at $\approx$ 4.1 V vs. Li$|$Li$^{+}$, which we used to intentionally condition the material with relevant voltage windows for this work. LMFP64 displays the classical redox picture due to an intense induction effect from the P ions. \cite{LIU2016109} In contrast, by carefully considering bond-disproportionate site, intersite charge fluctuation, and metal-ligand orbital hybridisation effects, a different redox view emerges in LNO more linked to a local site and ligand hole reconfiguration. Our findings provide critical insights to the electronic structure and a pathway to understanding oxygen redox.

\section*{RPES: Classical and Non-classical Redox}
\subsection*{The Classical Redox in LMFP64}
\begin{figure}[h]
  \includegraphics[width=\linewidth]{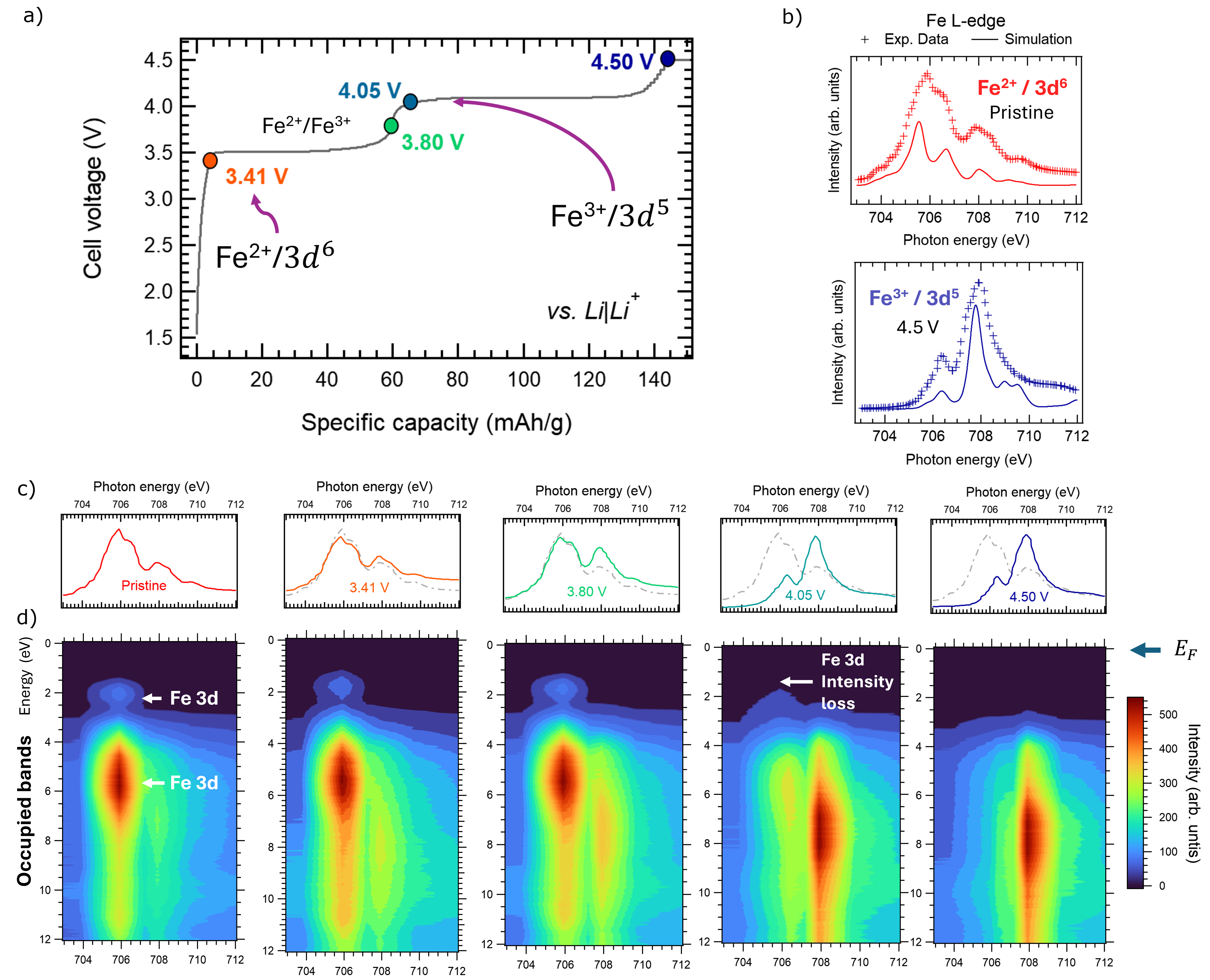}
  \caption{a) Electrochemical data of an LMFP64 half cell during C/20 charge with points indicating the cell potential points at which we disassembled the coin cells for XAS and RPES studies. b) Comparison of simulated spectra of Fe L-edge XAS with experimental data. c) Fe L-edge XAS measurements in the TEY mode of the LMFP64 electrodes at 3.41, 3.80, 4.05, and 4.50 V along with pristine electrodes to identify X-ray photon energies for the RPES studies. d) Heat maps of the RPES experiments showing the evolution of the electronic structure as a function of delithiation in LMFP64. Binding energy on the Y-axis of the RPES maps. The Fermi level $E_{F}$ is at zero binding energy mark and is highlighted with a blue arrow. c) and d) share the same x-axis.}
  %Reproduced with permission.\textsuperscript{[Ref.]} Copyright Year, Publisher. 
  \label{fig:LMFP_ResPES}
\end{figure} 

The Fe redox mechanism in LMFP64 represents the archetypal conventional model, where delithiation corresponds to electron depopulation from Fe \textit{3d} orbitals. Within our approach, we use this material as a reference to establish the spectral fingerprints of the conventional redox picture. To this end, we prepared a set of LMFP64 half-coin cell formats to charge the cells to targeted cell potentials as indicated by the coloured circles in Figure \ref{fig:LMFP_ResPES}a. In general, LMFP64 undergoes unexpected structural transformations under fast charge rates, \cite{Bak2018} which can affect intercalation reactions and, consequently, the rate of Fe-ion charge compensation. \cite{Bree2025} To minimise these effects, we employed a slow C/20 charging rate (7.93 mA/g), as described in the Methods section. This guarantees a cell potential profile with negligible polarisation and, as a result, a better approximation to the open circuit voltage profile (the true cell potential profile of cathode materials). \cite{Zhang2022} 

The top panel in Figure \ref{fig:LMFP_ResPES}b shows the experimental Fe L-edge spectrum of the pristine electrode (no cycling and no exposure to electrolyte) and the SIA simulated Fe L-edge spectrum. We performed the Fe L-edge simulations without explicit Fe \textit{3d} - O \textit{2p} ligand hybridisation (to guarantee an ionic picture) and with six Fe \textit{3d} electrons (to generate the expected 2+ valence). The remarkable match between the Fe L-edge experimental and simulated spectra confirms the ionic picture best describes the LMFP64 electronic structure; this means that an increase in the oxidation state inherently implies a reduction in the electron occupancy of the Fe \textit{3d} orbital. To simulate the fully charged condition at 4.5 V, we deliberately reduced the number of Fe \textit{3d} electrons from six to five in the simulation while keeping the no Fe \textit{3d} - O \textit{2p} ligand hybridisation condition. Spectral comparison shows again a remarkable match of the simulated and experimental Fe L-edge spectra (bottom panel in Figure \ref{fig:LMFP_ResPES}b). Combining these findings for the pristine and 4.5 V electrodes strongly suggests that the Fe redox mechanism involves a reduction in the number of Fe \textit{3d} electrons, leading to an increase in the Fe oxidation state from 2+ to 3+ in LMFP64 during charge. Notice that in the 704 - 709 eV energy range, Fe \textit{3d}-orbitals are the dominant feature of the Fe L-edge XAS transitions during charge (Figure~\ref{fig:LMFP_ResPES}c). Resonating in this range with RPES experiments will reveal the LMFP64 redox mechanism as a depopulation of the Fe \textit{3d} orbitals while the cell charges, proving the classical redox view in LMFP64.

Figure~\ref{fig:LMFP_ResPES}d shows RPES results at progressively increasing cell potential (pristine, 3.41, 3.80, 4.05, and 4.50 V, respectively) in the Fe \textit{3d} dominated energy range. With increasing potentials, we observe the RPES Fe \textit{3d} intensity at $\approx$ 2 eV binding energy decreases after the Fe$^{2+/3+}$ potential plateaus in Figure \ref{fig:LMFP_ResPES}a, experimentally confirming the conventional redox picture and that Fe oxidation state changes in LMFP64 links to the valence-counting formalism on the Fe \textit{3d} orbital. A similar procedure was employed to investigate the participation of the Mn sites in the redox mechanism. The analysis prompted further investigation of a possible degradation mechanism stemming from the Mn$^{2+/3+}$ redox couple that we will explore in a subsequent more in-depth manuscript regarding LMFP64 degradation.

\subsection*{LNO: Classical Redox?}
\begin{figure}[H]
\begin{center}
  \includegraphics[width=\linewidth]{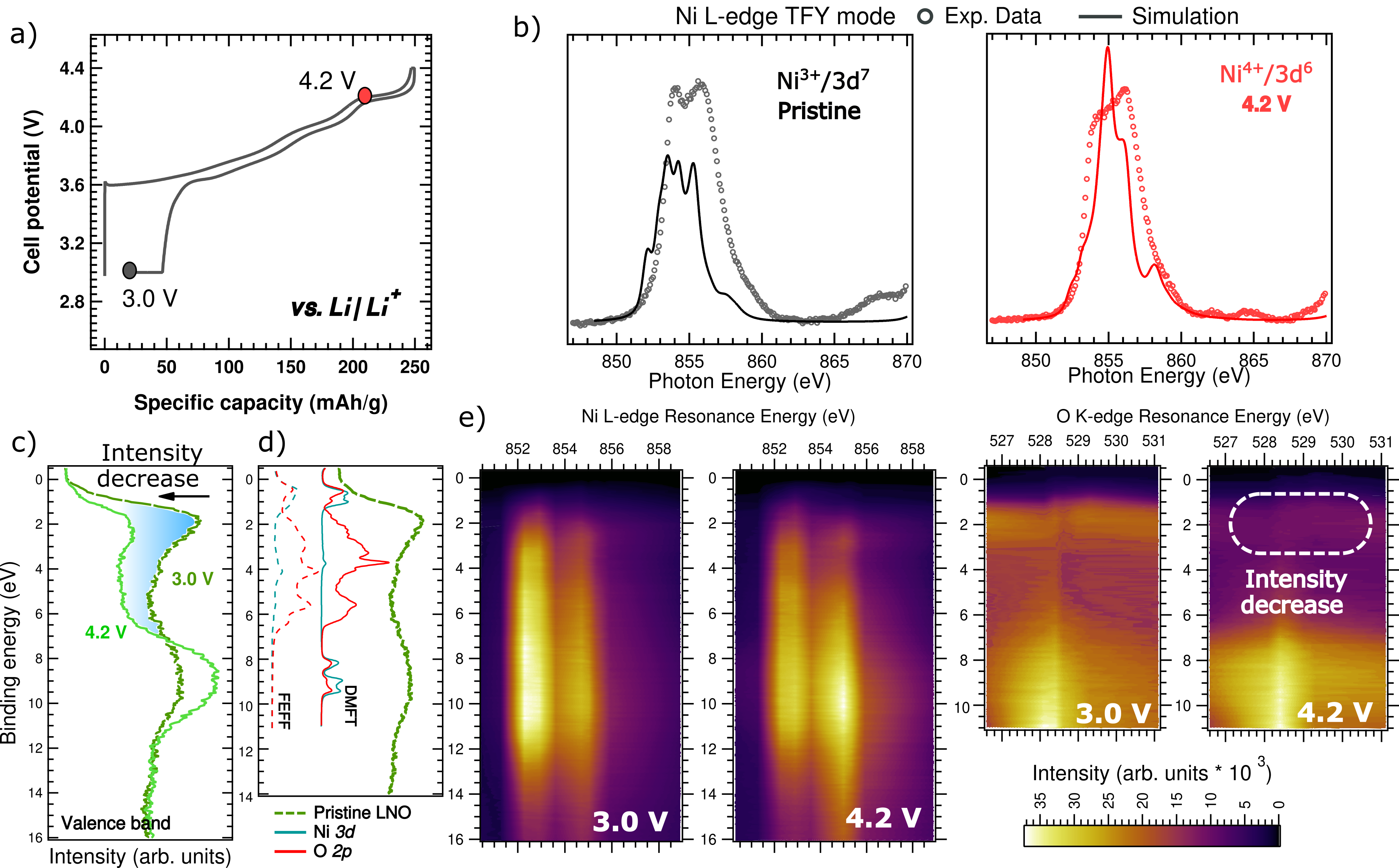}
  \caption{a) Electrochemical data of an LNO half cell during C/20 charge with points indicating the cell potential points at which we disassembled the coin cells for XAS and RPES studies. b) Comparison of simulated spectra of Ni L-edge XAS with experimental data in the TFY mode. c) XPS measurements of the valence band region for electrodes at 4.2 V (charged) and after discharge to 3.0 V, showing intensity changes that correlate with the corresponding Ni oxidation states at each voltage. d) DMFT and GW0-based calculations of the valence band density of states, identifying the spectral regions dominated by Ni \textit{3d} and O \textit{2p} character. d) RPES at the Ni L-edge, indicating minimal spectral change, consistent with negligible Ni \textit{3d} participation during redox, and RPES at the O K-edge, showing a significant loss of spectral intensity upon charge, attributed to oxidation and depletion of O \textit{2p}-derived ligand-hole states.}
  %Reproduced with permission.\textsuperscript{[Ref.]} Copyright Year, Publisher. 
  \label{fig:LNO_LigandHole}
\end{center}
\end{figure}

If Ni \textit{3d} valence-counting governs the redox mechanism in LNO cells, Ni L-edge RPES experiments should reveal spectral evolution similar to that observed for Fe. To test this hypothesis, we assembled half coin cells (vs. Li$|$Li$^{+}$) and charged them at C/20 (C=220 mA/g as described in the Methods section) to 4.2 V and later discharged them at 3.0 V, Fig. \ref{fig:LNO_LigandHole}a.

Contrary to the Fe case in LMFP64, Ni L-edge simulations in the ionic picture (i.e., without explicit Ni \textit{3d}-O \textit{2p} hybridisation) show poor agreement with experiment. Both the pristine state, simulated with 7 electrons in the Ni \textit{3d} orbital, and the charged state, simulated with 6 electrons, exhibit significant mismatch with bulk-sensitive Ni L-edge spectra in TFY mode (Fig. \ref{fig:LNO_LigandHole}b). This indicates that more complex electronic structures govern LNO redox during cycling, with electron depopulation arising from states beyond the simple Ni \textit{3d} orbital.

This more complex redox evolution is revealed through a strategic integration of RPES and theoretical frameworks, representing---to our knowledge---the first direct evidence of metal-ligand redox. In Fig. \ref{fig:LNO_LigandHole}c, we present valence-band XPS spectra of LNO electrodes at 4.2~V and after discharge to 3.0~V. While changes in the topmost valence band clearly track the electrochemical state, these changes alone lack both element specificity and orbital resolution. To resolve the orbital character, we employed DMFT and GW0 calculations (Fig. \ref{fig:LNO_LigandHole}d) alongside Ni L-edge and O K-edge RPES measurements (Fig. \ref{fig:LNO_LigandHole}e). Despite their distinct theoretical frameworks, both DMFT and GW0 consistently assign the valence-band edge primarily to a mixed Ni \textit{3d}-O \textit{2p} character, with the Ni \textit{3d} $t_{2g}$ states positioned deeper, around 10 eV. This assignment suggests that LNO redox involves spectral evolution within this strongly hybridised Ni \textit{3d}-O \textit{2p} orbital. 

RPES measurements directly confirm the theoretical assignment of Ni \textit{3d}-O \textit{2p} hybridisation and substantiate its role in governing LNO redox. Ni L-edge RPES shows an intense, electrochemically invariant feature at $\sim$10~eV, consistent with theory predicting deeply bound Ni \textit{3d} states. These observations strongly indicate that the dominant Ni \textit{3d} electron occupancy remains unchanged during cycling, i.e., throughout LNO redox. When resonating at the O K-edge pre-edge, the RPES spectrum reveals a strong feature at the same $\sim$10 eV, confirming substantial Ni \textit{3d}-O \textit{2p} hybridisation. Crucially, only the O K-edge RPES exhibits clear discharge-induced changes, particularly in the top valence states. Building on the Fe case, where such changes signify electron occupancy variation, the combined Ni L-edge and O K-edge RPES trends indicate that LNO redox fundamentally involves electron population changes on ligand states in strongly hybridised orbitals.

Collectively, our results reveal an intriguing shift in the redox mechanism, from metal-centred in LMFP64 to ligand-dominated through cooperative Ni-O hybridisation, and thereby open new avenues of inquiry: why does this shift occur, what drives it, and how does it shape our broader understanding of redox chemistry?

\section*{Electronic Structures: From Metal-Centred to Hybridised} \label{sec;fingerprints}

The shift in the redox mechanism described in the previous section arises from electronic structures, which evolve toward greater complexity with increasing formal oxidation state of the metals. At higher oxidation states, the stronger electron affinities of the metals and the weaker electronegativities of the ligands combine to produce compounds with more intricate electronic structures, \cite{Pavarini:819465} thereby defining the pathways through which redox occurs. Using our approach, we designed experiments to identify the spectroscopic signatures of this electronic evolution and lay the foundation to bridge fundamental insights with cell-level redox performance, establishing a reliable from-quantum-to-cell loop for a comprehensive understanding of LNO redox in the next section. 

\begin{figure}[h]
\begin{center}
  \includegraphics[width=\linewidth]{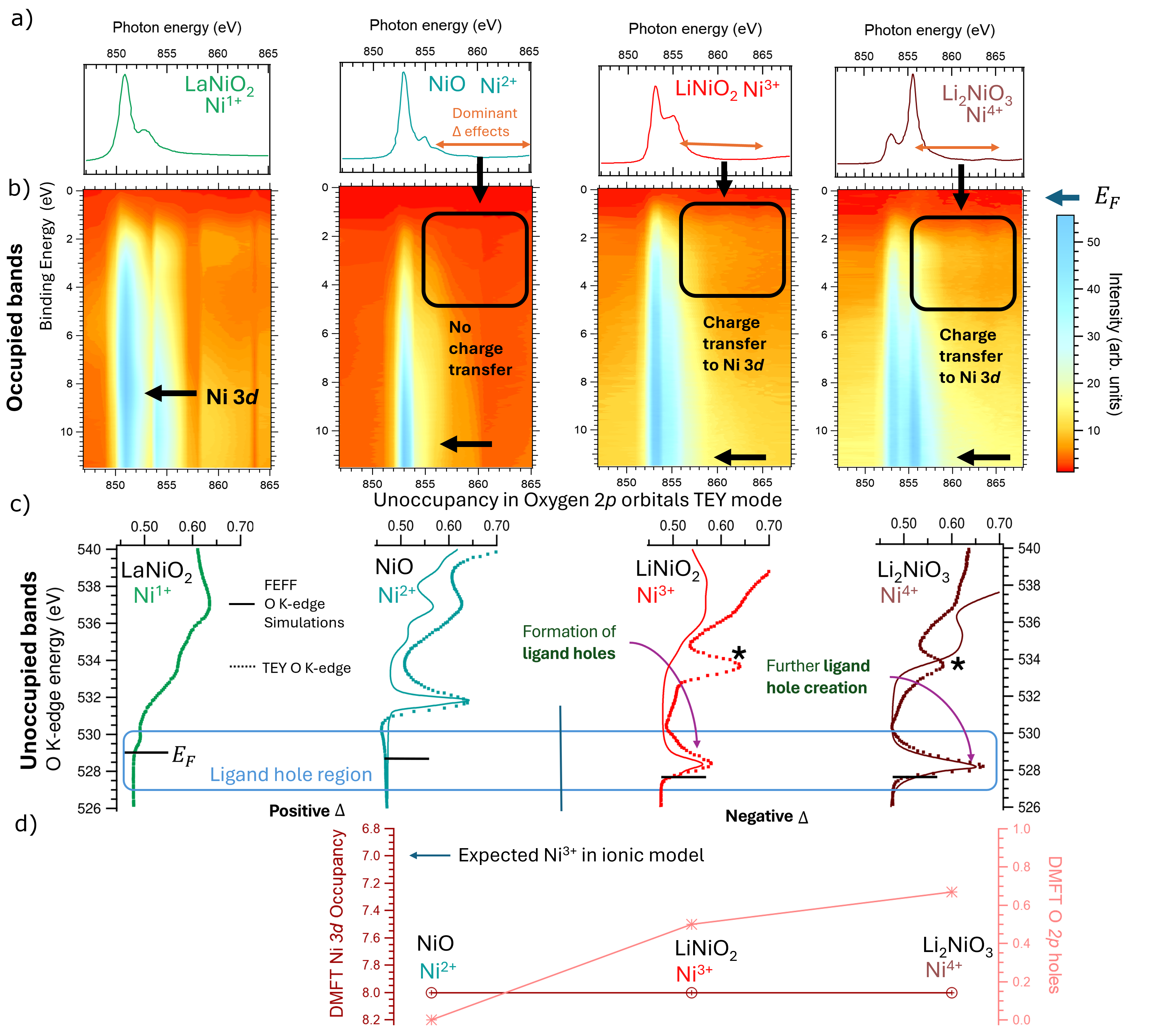}
  \caption{a) Ni L-edge XAS measurements in the TEY mode of various Ni oxide compounds to identify X-ray photon energies for the RPES studies. b) Heat map showing the intensities of the RPES experiments indicating the energies of occupied orbitals and the effects of charge transfer into the Ni 3\textit{d} from the ligand orbital. c) Unoccupied O \textit{2p} orbitals in TEY mode O K-edge experiments (dotted lines) and Green's function-based O K-edge simulations (solid lines) with a peak whose intensity increases due to ligand hole formation as a result of spontaneous charge transfer from the O \textit{2p} ligand to Ni \textit{3d} $e_{g}$ in the negative $\Delta$ regime. *typical carbonate (Li$_{2}$CO$_{3}$) surface contamination only affecting high-valence Ni oxides. d) DMFT-derived occupancies of the Ni \textit{3d} orbital and the O \textit{2p} orbital correlating with the increasing pre-edge peak in O K-edge as a combined signature of ligand-hole creation dominating the oxidation state evolution of Ni oxides}
  %Reproduced with permission.\textsuperscript{[Ref.]} Copyright Year, Publisher. 
  \label{fig:NiOxides_NCT}
\end{center}
\end{figure}

To this purpose, we prepared a set of four Ni-oxide compounds (LaNiO$_2$, NiO, LNO and Li$_2$NiO$_3$), where the Ni ion is oxidised in different formal oxidation states (i.e$.$ Ni$^{1+}$, Ni$^{2+}$, Ni$^{3+}$ and Ni$^{4+}$). The Ni-O ligand hybridisation is minimal in LaNiO$_2$ \cite{Hepting2020} while playing a substantial role in NiO. \cite{PhysRevB.100.115125} LNO is better described with intersite charge fluctuations, explicit TM \textit{3d} - O \textit{2p} ligand hybridisation, and bond-disproportionate NiO$_{6}$ sites. \cite{depth_resolvingACSenergy} Li$_2$NiO$_3$ is the endpoint of Li-excess and its electronic structure is better described by intersite charge fluctuations and explicit TM \textit{3d} - O \textit{2p} ligand hybridisation. \cite{Electrochemical_testing_li2nio3} Therefore, our approach applied to the Ni oxide material set will show the spectral difference of the transition towards more complex electronic structures.

Figure~\ref{fig:NiOxides_NCT}a shows Ni L-edge XAS spectra of as-synthesised LaNiO$_2$, NiO, LNO, and Li$_2$NiO$_3$ ,with the corresponding RPES measurements in Figure~\ref{fig:NiOxides_NCT}b. Given the shallow probing depth of RPES measurements, Ni oxides prone to surface reconstruction can produce spectral convolution between surface species and the subsurface region. Figure S1 identifies that surface spectral contribution concentrates at $\approx$ 853 eV and energies $>$ 854 eV are redox-related originating from subsurface regions beneath the surface-reconstructed layer. We thus focused our RPES experiments and analysis on Ni L-edge transitions with energies $>$ 854 eV.

Building on the LMFP64 case, where the redox-related spectral signature appears at the topmost rising edge, our analysis shows that the Ni \textit{3d} $t_{2g}$ bands (indicated by arrows in Figure~\ref{fig:NiOxides_NCT}b) are not redox-active, as they lie deeper in energy (8-10 eV) across the Ni-based oxide set, consistent with the LNO cathode observations discussed in the previous section. Instead, the states located near E$_{F}$---within the redox-active regions, highlighted by the white square in Figure~\ref{fig:NiOxides_NCT}b---are redox-relevant. Resolving the nature of these states, by identifying XAS transitions through which they resonate in Ni L-edge RPES (855-865 eV), is key to understanding the evolution from metal-centred to hybridised electronic structures and the associated redox processes in Ni compounds. By combining dynamical mean-field theory (DMFT) for orbital occupancy analysis, Ni L-edge transition simulations within SIA models, and O K-edge calculations using real-space multiple-scattering theory in an interacting Green's function basis (FEFF), we provide strong multi-level evidence that spontaneous ligand-to-metal charge transfer drives the electronic structure evolution, thereby defining the redox-active bands that ultimately dictate the redox pathways during Ni-based cell operation.

As the formal oxidation state increases, the cooperative effect of stronger electron affinity and weaker electronegativity reaches a threshold at which, in strongly hybridised orbitals, charge redistributes between Ni and O, allowing partial charge transfer---also referred to as charge fluctuation---to stabilise. Experimentally, this phenomenon is parametrised by a factor $\Delta$, with positive values indicating the stage prior to, and negative values marking the onset of, spontaneous ligand-to-metal charge transfer. As demonstrated by SIA-based Ni L-edge simulations of the positive-$\Delta$ NiO system (Fig. S2), these charge fluctuations dominate transitions in the Ni L-edge XAS energy range (855-865 eV), where resonance in RPES gives rise to the Ni-O hybridised orbitals at the top of the VB, i.e. the redox-relevant states of Ni compounds. Explicit inclusion of Ni \textit{3d}-O \textit{2p} ligand hybridisation in the simulations (Fig. S2b) was critical to resolving the features within this 855-865 eV range. However, because $\Delta$ remains positive for NiO, this fluctuation is not spontaneous but instead arises from X-ray excitation---commonly referred to as the Zhang-Rice bound state \cite{PhysRevB.100.115125}---and is therefore not intrinsic to NiO's electronic structure, explaining why NiO lacks spectral features of charge-transfer-related states in the Ni L-edge RPES resonance between 855-865 eV. We therefore label the 855-865 eV window as the $\Delta$-dominant region, since its RPES resonance links directly to hybridised orbitals with spontaneous charge-transfer character, which are expected to be prominent in negative-$\Delta$ systems.

Indeed, the pronounced Ni L-edge RPES features resonating in this $\Delta$-dominant range for LNO and Li$_{2}$NiO$_{3}$ indicate that the transition to the negative-$\Delta$ regime occurs between Ni$^{2+}$ and Ni$^{3+}$. Thin films prepared to simulate this Ni oxidation progression further support this view, implying expectations of non-conventional redox in Ni$^{3+}$ systems arising from their negative-$\Delta$ character. DMFT orbital occupancy analysis (Fig. S3) confirms this spontaneous charge migration beyond Ni$^{2+}$, stabilised through the gradual formation of electronic ``holes'' or $\underline{L}$ in the O \textit{2p} ligand state within the hybridised orbitals. Spectral measurements (Fig. S3) validate these theoretical predictions, revealing hole formation as intensity migration from the topmost occupied states (hybridised Ni \textit{3d}-O \textit{2p} ligand) to the lowest unoccupied states (O K-edge pre-peak). The rising O K-edge pre-peak thus serves as a clear manifestation of ligand hole formation and stabilisation in the hybridised Ni \textit{3d}-O \textit{2p} ligand orbital, consistent with previous reports. \cite{PhysRevB.45.1612, PhysRevLett.62.221}

Following this interpretation for Fig.~\ref{fig:NiOxides_NCT}b-d, LNO and Li$_{2}$NiO$_{3}$ exhibit clear signatures of negative-$\Delta$ systems. Their formal oxidation states exceed Ni$^{2+}$, placing them well beyond the positive-to-negative $\Delta$ transition threshold. This results in stabilised ligand holes, as confirmed by DMFT calculations in both compounds. Experimentally, the presence of ligand holes manifests as pronounced Ni L-edge intensity in the $\Delta$-dominated range. The progressive growth of the O K-edge pre-edge feature also reflects the DMFT-predicted increase in ligand-hole numbers as oxidation proceeds from Ni$^{3+}$ to Ni$^{4+}$, consistent with the corresponding growth of the pre-edge peak in FEFF-based simulations. Together, these observations establish the O K-edge pre-edge as a robust descriptor of ligand holes, endorsing ligand-hole formation as the mechanism driving evolution toward higher Ni oxidation states. Crucially, the strong consistency in describing ligand-hole formation across three distinct theoretical frameworks---DMFT, multiplet effects in SIA, and real-space correlation in GW0 (FEFF)---reconciles these approaches and underscores both the robustness of this description and the central role of ligand holes in dictating the redox pathways of high Ni oxidation-state systems.

This integrated experimental and theoretical analysis firmly establishes LNO and Li$_{2}$NiO$_{3}$ as prototypical negative-$\Delta$ systems, in which ligand-hole is already present in the pristine state and its formation and annihilation is the dominant mechanism governing the evolution across the full $\text{Ni}^{2+} \leftrightarrow \text{Ni}^{3+} \leftrightarrow \text{Ni}^{4+}$ range.

It is worth noting that a hole in the ligand orbital is a source of instability which a positive charge transfer system often corrects, annihilating the hole, in time intervals of femtosecond via radiation or thermal vibration. However, in negative charge transfer materials, the electronic structure modifications near $E_{F}$ manifest as a correlated redistribution of charge, leading to various newly-formed hybridised states where the hole coexists with the compensating Ni \textit{3d} charge.

\section*{Understanding Redox in a Ligand Hole Material} \label{sec:RedoxLigandHole}
The analysis presented in previous \hyperref[sec;fingerprints]{section} provides a robust framework to investigate the redox process in LNO positive electrodes. In Fig. S4, as the cell potential increases, the O K-edge pre-edge feature, previously associated with ligand hole formation, exhibits a concurrent intensity growth (Fig. S4b). This trend suggests that since the pristine material already exhibits this pre-edge spectral feature, the electrochemical charging process, across the full voltage range, is governed by the progressive formation of ligand holes. This mirrors the systematic evolution of the O K-edge pre-edge observed across the Ni oxide series in Figure~\ref{fig:NiOxides_NCT}c. Resonance in the $\Delta$-dominated range in Ni L-edge RPES measurements (Fig.~S4c) reveals a spectral attenuation of the ligand-hole hybridised orbitals as cell potential increases. This reduction, highlighted in Fig. S4d (white squares), indicates a progressive electron withdrawal from the ligand-hole states, confirming their direct participation in LNO charge compensation.

\begin{figure}[H]
\begin{center}
  \includegraphics[width=\linewidth]{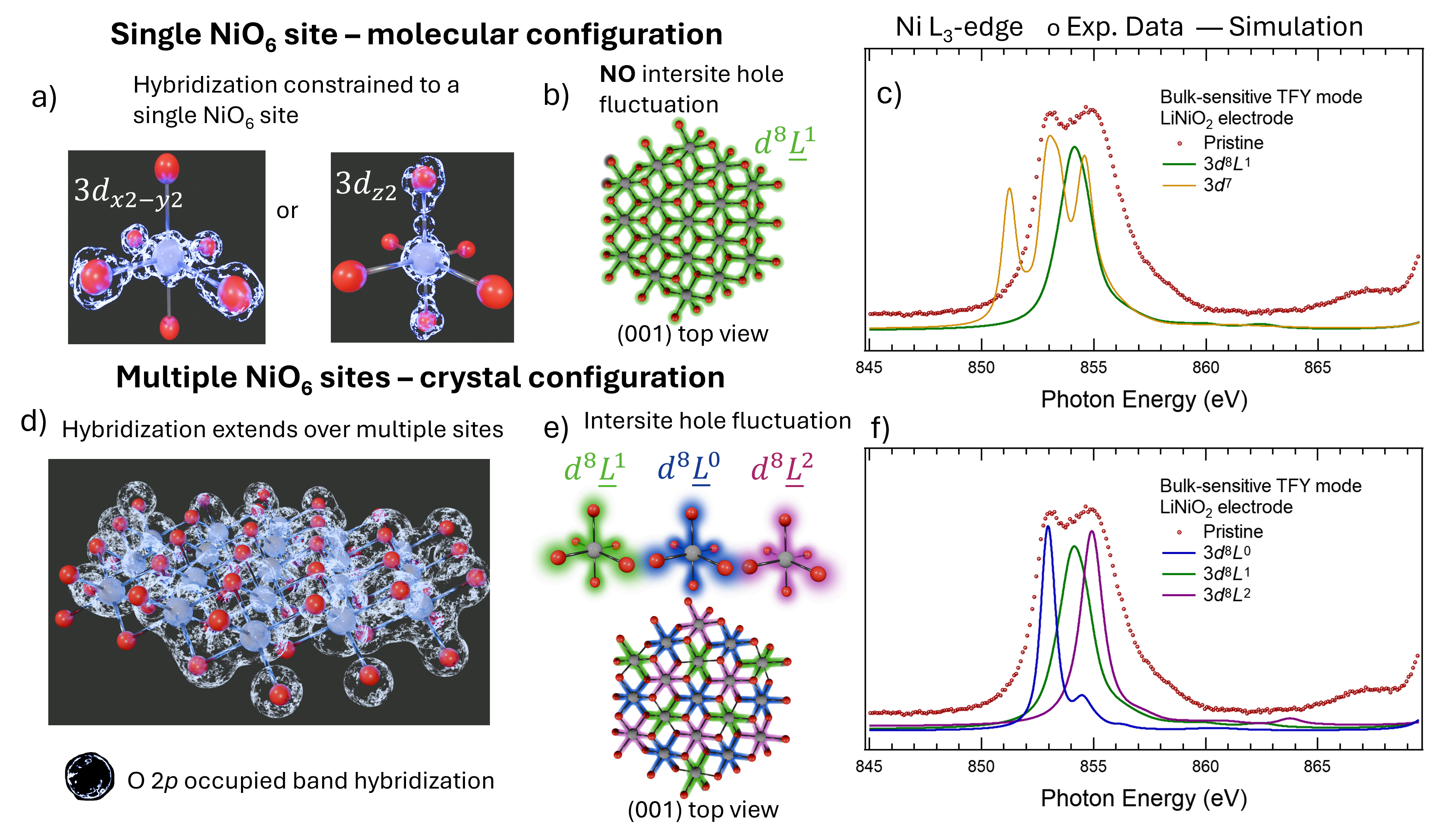}
  \caption{The space occupied by the hybridised Ni \textit{3d} - O \textit{2p} ligand orbital relative to a) a single NiO$_{6}$ site or d) over multiple sites has an impact on the type of hole configuration (b and e) forming on pristine LNO electrodes. Comparing simulations with bulk sensitive Ni L-edge XAS experiments (c and f) clarifies the true type of ligand hole configuration in pristine LNO.}
  %Reproduced with permission.\textsuperscript{[Ref.]} Copyright Year, Publisher. 
  \label{fig:LNO_MultipleHoles}
\end{center}
\end{figure}

The specific nature of these ligand-hole states remains to be directly determined for a comprehensive description of LNO under this formalism. To address this, we now focus on bulk-sensitive Ni L-edge measurements of the pristine LNO electrode, in conjunction with Ni L-edge simulations in the relevant $\Delta$ regime. In Figure \ref{fig:LNO_MultipleHoles}, we start with an electronic ground state assumed to have a fully Ni$^{3+}$ configuration (3d$^{7}$) with explicit hybridisation. This case illustrates the influence on the electronic structure of hybridisation alone without spontaneous ligand hole stabilisation. The significant mismatch between the resulting simulated Ni L-edge spectrum and the experimental one further confirms the relevance of the ligand hole formalism over the simple 3d$^{7}$ model.  We continue with considering $d^{8}\underline{L}^{1}$ as the main ligand hole configuration in pristine LNO. When one reads this configuration in the literature, a couple of preconditions are often overlooked. First, the spatial extent of this hybridised orbital is assumed to be confined to a single NiO$_{6}$ site. And second, due to this single-site spatial extension, the $d^{8}\underline{L}^{1}$ state results from the hybridisation between the Ni \textit{3d} \textit{e$_{g}$} orbitals and the O ligand orbitals that directly point to the Ni ion as seen in Figure \ref{fig:LNO_MultipleHoles}a. These conditions lead to a NiO$_{6}$ layer where $d^{8}\underline{L}^{1}$ ligand configuration is adopted homogeneously by every individual octahedra, as seen in Figure \ref{fig:LNO_MultipleHoles}b. Figure \ref{fig:LNO_MultipleHoles}c shows that theoretical calculations based on such a single ligand holes configuration reproduce poorly the measured Ni L$_{3}$-edge absorption spectrum because these theoretical conditions don't consider the entire crystallographic structure.

A more realistic approach is considering that the orbital hybridisation expands across the entire structure as in the case of band theory. Indeed, when the Ni $3d$-O $2p$ ligand hybridisation extends across multiple NiO$_6$ octahedra, ligand holes can be more easily stabilised via a ``self-regulating response''. \cite{PhysRevB.98.075135} This indicates that transferring ligand electrons to the metal is regulated collectively within a set of neighbouring NiO$_{6}$ octahedra in which the ligand holes reorganise among them via ligand-metal bond alternation, leading to configurations of the type seen in Figure \ref{fig:LNO_MultipleHoles}e. These configurations have been observed in structural experiments of LNO. \cite{PhysRevB.100.165104, PhysRevB.109.035139} As a result, we have ligand configurations of the type $d^{8}\underline{L}^{0}$, $d^{8}\underline{L}^{1}$, and $d^{8}\underline{L}^{2}$ that correspond geometrically to unique octahedral configurations. Our hypothesis of the LNO electronic configuration having an equal distribution of these three types replicates well the XAS spectra, as shown in Figure~\ref{fig:LNO_MultipleHoles}f, confirming the self-regulating response is a realistic electronic structure for the pristine LNO electrode. Here, we simulated the Ni L-edge transitions with the three electronic configurations in three independent SIA models. Although the highly covalent yet correlated nature of LNO requires a more complex single Ni L-edge simulation with multiple clusters of NiO$_{6}$ to properly capture bond-alternation, correlation, and covalency at once, \cite{PhysRevB.94.195127, Green_2020_bonddisproportionation} the independent SIA approach herein adopted provides key insights into the basic nature of the Ni L-edge transitions and the corresponding experimental peaks.

\begin{figure}[H]
\begin{center}
  \includegraphics[width=\linewidth]{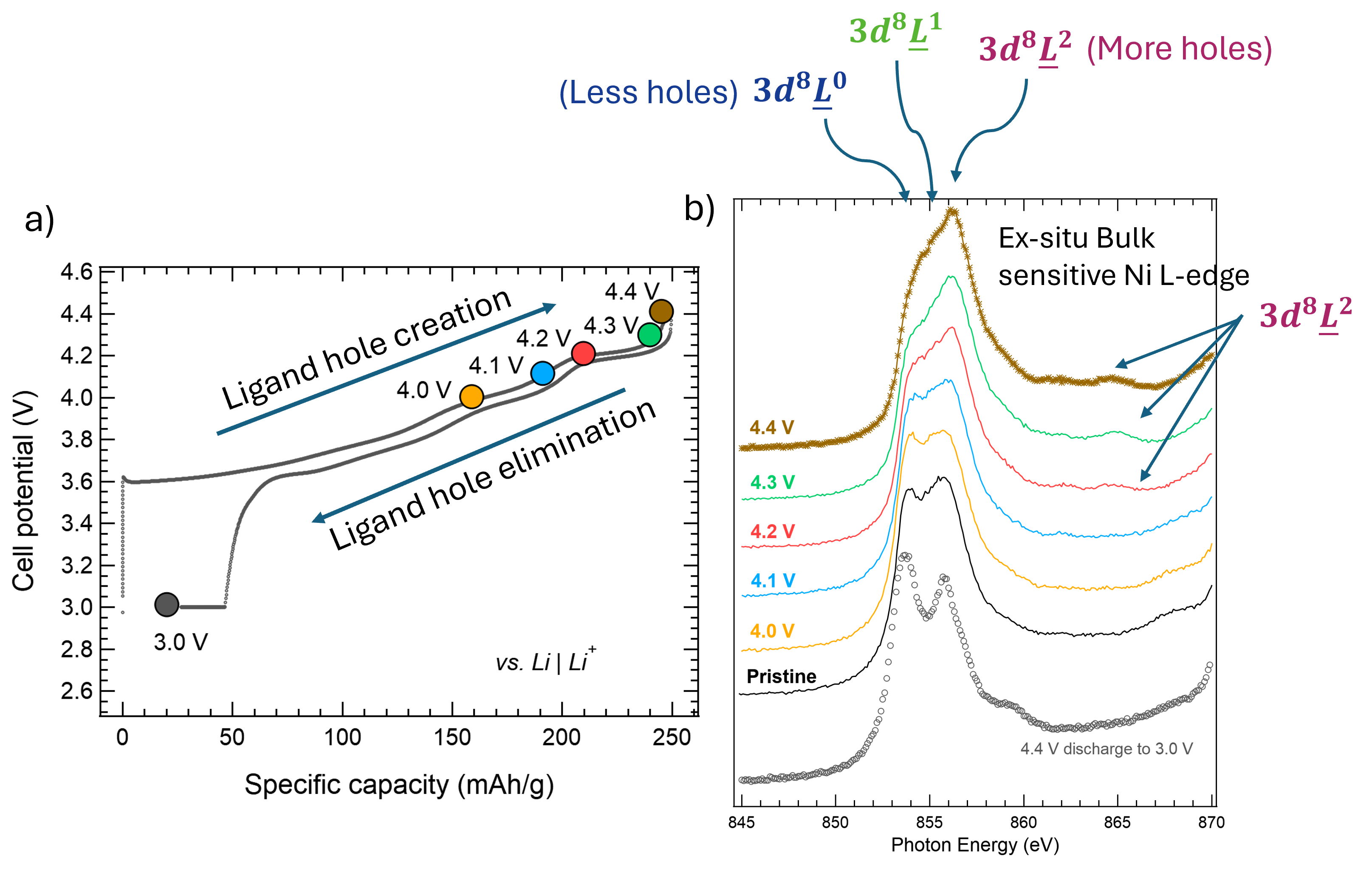}
  \caption{a) Electrochemical data of LNO half cells during a C/20 charge protocol with marks indicating cell potential points corresponding to cell disassembly for ex-situ TFY mode Ni L-edge XAS studies. b) Ni L-edge in the TFY mode (bulk sensitive) of LNO electrodes retrieved from cells disassembled at the given cell potentials.}
  %Reproduced with permission.\textsuperscript{[Ref.]} Copyright Year, Publisher. 
  \label{fig:LNO_HolesCreationRedox}
\end{center}
\end{figure}

Understanding the redox mechanism at the LNO positive electrode side during the charging process ultimately comes down to clarifying how the three $d^{8}\underline{L}^{0}$, $d^{8}\underline{L}^{1}$, and $d^{8}\underline{L}^{2}$ ligand configurations adjust during intercalation reactions. In this sense, we have run bulk sensitive Ni L-edge XAS measurements at different cell potentials along the electrochemical profile of half coin cells (also charged at a C-rate of C/20), which are highlighted in Figure~\ref{fig:LNO_HolesCreationRedox}a. From our previous calculations in Figure~\ref{fig:LNO_MultipleHoles}f we found that the XAS spectra can be interpreted in terms of the intensity of three peaks (at roughly 854, 855 and 856~eV), which denote the LNO crystallographic structures of three different types of NiO$_{6}$ octahedra and ligand hole configurations. As the battery is charged up to 4.4~V, there is a gradual increase in the intensity of the $d^{8}\underline{L}^{2}$ peak, while the $d^{8}\underline{L}^{0}$ peak is reduced. This suggests that LNO cell charging involves creating new holes that the $d^{8}\underline{L}^{0}$ configuration absorbs. As a result, NiO$_{6}$ sites reconfigure, Ni-O ligand re-hybridise, and charge redistribute leading to more instances of the $d^{8}\underline{L}^{1}$, and $d^{8}\underline{L}^{2}$ configurations in alignment with the data in Figure~\ref{fig:LNO_LigandHole}. This spectral evolution concentrates in the $\Delta$-dominated Ni L-edge range which in tandem reflects the DMFT-predicted increase of ligand holes, reconciling these fundamentally distinct theoretical pictures used to describe ligand hole evolution. As expected, the discharge process reverses the spectral evolution of the charge process. The filling of these ligand holes dominates the discharge process which we can see in the spectral reduction of the $d^{8}\underline{L}^{2}$ and the increasing spectral features of the $d^{8}\underline{L}^{0}$ configuration after discharging from 4.4 V to 3.0 V vs. Li$|$Li$^{+}$. A direct consequence of the hole creation/filling dynamics is that electrochemical cycling modulates the distribution of local structures, specifically NiO$_{6}$ octahedra. We anticipate a coexistence of distinct octahedral geometries associated with the $d^{8}\underline{L}^{0}$, $d^{8}\underline{L}^{1}$, and $d^{8}\underline{L}^{2}$ ligand-hole states. A follow-up study will investigate this local structural distribution using operando Ni K-edge XAS to elucidate its dependence on hole formation and (de)intercalation reactions in stoichiometric Ni-rich materials. Instead of counting Ni \textit{d} electrons, in negative charge transfer compounds, the redox mechanism refers to counting those hole-induced distorted octahedral sites, caused by the dynamical evolution of ligand electronic density.

\section*{Redox Mechanisms and the ZSA Scheme}\label{sec:Discussion}

\begin{figure}
  \includegraphics[width=\linewidth]{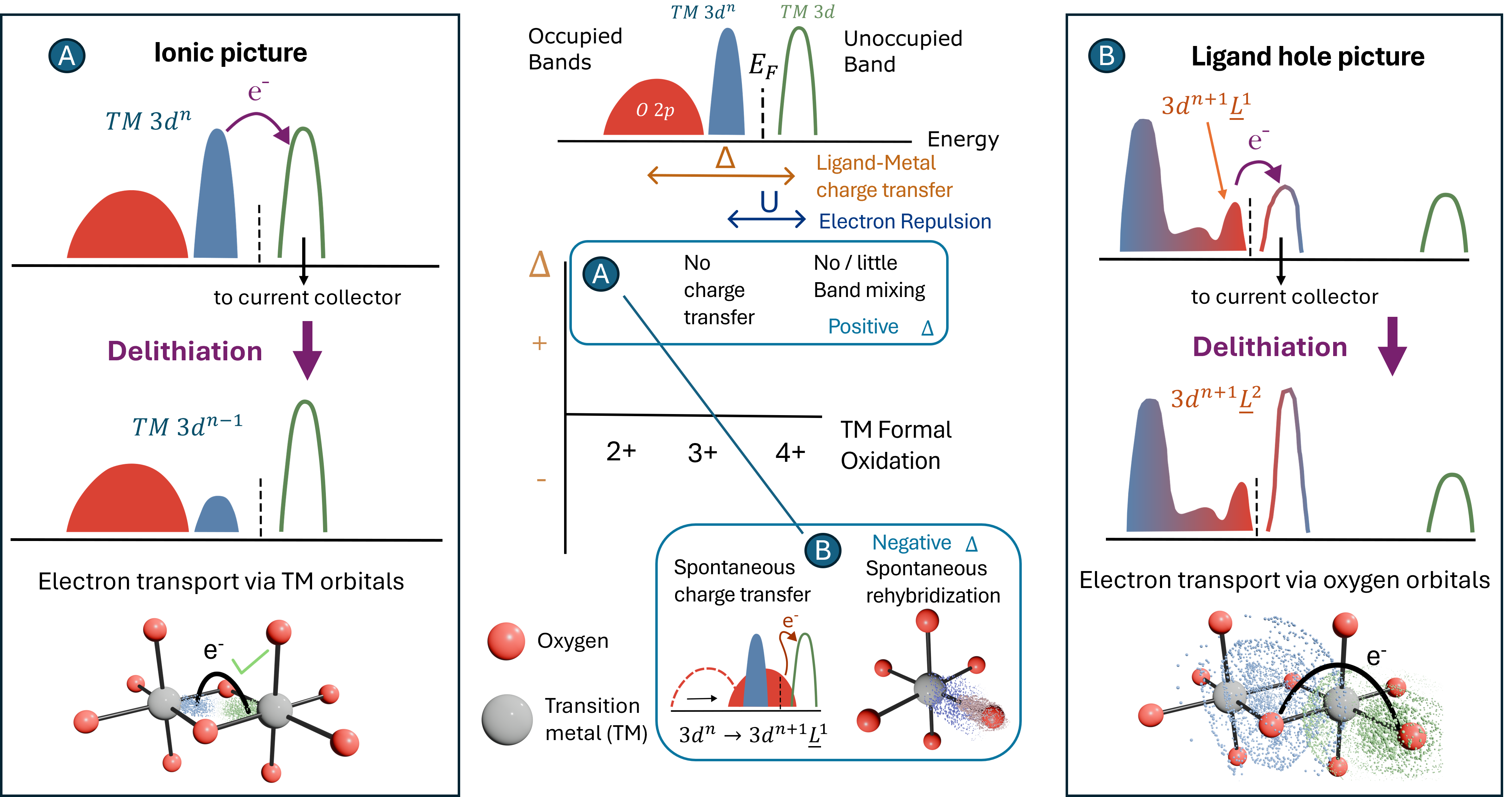}
  \caption{Schematics of the two end-points ZSA electronic models determined as a function of the relative strength of U and $\Delta$. The ionic picture (model A) is the ZSA Mott-Hubbard material class with $\Delta$ positive and $\Delta$ $>$ U. The ligand hole picture (model B) is the ZSA Negative Charge Transfer material class with U $>$ $\Delta$, with U positive and $\Delta$ $<$ 0. Type of gap between the conduction and valence band in these models dictates the type of redox process during (de-)intercalation reactions. The way we understand oxidation state evolution is then inherently linked to the relative strength of U and $\Delta$.}
  %Reproduced with permission.\textsuperscript{[Ref.]} Copyright Year, Publisher. 
  \label{fig:electronicModels}
\end{figure}

The precise description of the electronic structure of LMFP64 and LNO has been critical for the distinct charge compensation pathways for these materials. We inferred their electronic structures using the Zaanen-Sawatzky-Allen (ZSA) electronic scheme. It was first discussed by Zaanen, Sawatzky and Allen, \cite{PhysRevLett.55.418} where the relative energy positions of the metal \textit{d} and oxygen \textit{p} bands was used to categorise correlated transition metal compounds; a more recent work \cite{SZA_2024_negativeCHtransfer} defines four categories based on the material's relative energy values between $\Delta$ and U, which is the on-site Coulomb interaction, i.e., the electron repulsion causing a separation in energy of the \textit{3d}-derived electronic states. $\Delta$ is defined similarly as in the RPES analysis. The material band gap type formed within the ZSA electronic framework for LMFP64 and LNO dictates the corresponding charge compensation mechanism, illustrating how this band gap can aid in determining the expected material redox nature in general (Figure \ref{fig:electronicModels}).

Model A in Figure \ref{fig:electronicModels} describes the ZSA Mott-Hubbard material class---positive $\Delta$ ($\Delta$ $>$ U)---applicable to the LMFP64 case. The band gap results from Coulomb repulsion, so it is of a \textit{d} - \textit{d} nature. This means that the lowest unoccupied band (conduction band) and highest occupied band (valence band) are both of Fe \textit{d} character. During cell charge, charge from the Fe \textit{3d} valence band will be promoted to the conduction band. The loss of Fe intensity in Figure \ref{fig:LMFP_ResPES}d is a manifestation of this effect, hence, Fe oxidation. Model B in Figure \ref{fig:electronicModels} defines the ZSA Negative Charge Transfer material class which determines the LNO electronic structure. Since in these materials, $\Delta$ is negative and U positive, the band gap is of a \textit{dp} - \textit{dp} type, as shown by our DMFT/RPES analysis in Fig.~\ref{fig:LNO_LigandHole}, \ref{fig:NiOxides_NCT}, S2, and S4. During the charge process of LNO cells, electrons are promoted across this \textit{dp} - \textit{dp} band gap, in a similar fashion as the spectral migration in Fig. S3, creating holes in the hybridised ligand orbitals while drawing capacity. The LNO system instantaneously undergoes a correlated redistribution of charge resulting in the reconfiguration of the $d^{8}\underline{L}^{0}$, $d^{8}\underline{L}^{1}$, and $d^{8}\underline{L}^{2}$ states during electrochemistry. As result, the electrochemical redox of LNO (and related Ni-rich cathodes) can be described by ligand hole formation and annihilation, as shown with Fig.~\ref{fig:LNO_HolesCreationRedox}.

This ZSA-based redox framework has immediate implications for high-valence Co compounds, which, consistent with recent reports, \cite{LCO_ligand_D4CP03759F} are expected to exhibit ligand-hole characteristics for Co oxidation states above 3+. In conjunction with the Ni results presented here, this suggests that NMC positive electrodes must display a strongly state-of-charge-dependent ligand-hole contribution. Furthermore, the framework directly informs emerging design rules aimed at stabilising ligand holes in high-valence Fe compounds, again in agreement with recent findings. \cite{Ramachandran2024} Beyond these specific chemistries, it establishes a generalised redox strategy applicable to lithium-excess compounds---the systems that originally motivated the study of oxygen redox. Note that the transition from positive to negative charge transfer regimes depends on metal and its formal oxidation state, corresponding to roughly Ni$^{2+}$, Co$^{3+}$ and Mn$^{4+}$. \cite{SZA_2024_negativeCHtransfer} Moreover, the majority of Li-excess compounds are Li-rich Mn-rich. \cite{Zhang2022} Li$_{2}$MnO$_{3}$ and Li$_{2}$NiO$_{3}$ reflect the metal and ligand hole endpoints, respectively, consistent with the observed electrochemistry arising from structural and chemical decomposition. \cite{Zuba_ACSEL_li2MnO3, Electrochemical_testing_li2nio3} Thus, we anticipate these materials could be classed in one of the other two ZSA categories (Positive Charge Transfer or Mixed Valence) with varying \textit{p}-\textit{d} band gap types, \cite{Pavarini:819465} offering a pathway to study the role of unhybridised oxygen orbitals in delivering reversible capacity without conflicting signals. We acknowledge that structural rearrangements also play a key role in the electrochemical behaviour of Li-rich systems. \cite{Assat2018} TM L-edge assisted RPES offers an opportunity to directly explore electronic structures of Li-excess compounds under the ZSA scheme during (de)lithiation and to validate models and descriptions that clarify the role of oxygen oxidation in these compounds.

\section*{Conclusion}
Here, we present how by experimentally capturing the true nature of the electronic structure of LiMn$_{0.6}$Fe$_{0.4}$PO$_{4}$ and LiNiO$_{2}$ positive electrode materials under the ZSA scheme, the TM L-edge assisted RPES approach can directly probe their corresponding redox mechanism. Using this framework, we can conclude that the degree of metal-oxygen hybridisation is a key factor in determining the charge compensation mechanism. For more ionic compounds in the positive charge regime, such as the phosphates, electron counting of the \textit{d} orbital remains valid  (e.g. Fe$^{2+/3+}$ $\equiv$ \textit{3d}$^{6}$,\textit{3d}$^{5}$). Increasingly covalent systems in the negative charge transfer regime, which includes the LiMO$_{2}$ (M= Co, Ni) derived family, are better described in terms of electron counting ligand hole states instead (e.g. Ni$^{3+/4+}$ $\equiv$ $d^{8}\underline{L}^{1}$, $d^{8}\underline{L}^{2}$). We expect this framework to also help describe lithium excess compounds. A more accurate probe into the evolution of their electronic structure may resolve the true role of unhybridised oxygen orbitals in delivering reversible capacity in these systems.

\backmatter

\bmhead{Supplementary information}
The supplementary information includes: Surface versus bulk analysis to identify energy regions in Ni L-edge experiments without surface reconstruction contribution, comparison between Ni L-edge simulations in models with and without explicit orbital hybridisation, integrated experimental and computational data to showcase how Ni oxidation state modulate negative charge transfer energy $\Delta$, and RPES and electrochemical data supporting a ligand hole redox mechanism in LNO.
%If your article has accompanying supplementary file/s please state so here. 

%Authors reporting data from electrophoretic gels and blots should supply the full unprocessed scans for key as part of their Supplementary information. This may be requested by the editorial team/s if it is missing.

%Please refer to Journal-level guidance for any specific requirements.

\bmhead{Acknowledgements}

HB acknowledges support from the Royal Society of Chemistry Research Fund Grant R24-8964281967. We acknowledge Diamond Light Source for time on beamline B07 under proposal SI33459-1 and beamline I09 under proposals SI35075-1, SI36917-1, and SI30201. This work was supported by the Faraday Institution's Degradation (FIRG060, FIRG082) project and the FutureCat (FIRG065, FIRG017) project.

%\bmhead{Author Contributions}

\section*{Online Methods}\label{sec:Methods}

\subsection*{Ni-based Materials}
These four powder materials were used during the SI36917 RPES beamtime at the I09 beamline of Diamond Light Source (DLS). Material preparation was performed avoiding air-exposure following the description provided in the RPES section.
\subsubsection*{LaNiO$_{2}$}
Single crystals of LaNiO$_{3}$ in the perovskite phase were grown using the high-pressure optical floating zone method.\cite{Planck_10.1063/5.0160912} The crystals were oriented with x-ray Laue diffraction and cut into cube-shaped pieces with surface dimensions of approximately 1 mm$^{2}$. Subsequently, oxygen was deintercalated via a direct-contact topotactic reduction with CaH$_{2}$, transforming the perovskite phase into the LaNiO$_{2}$ infinite-layer phase.\cite{Planck_PhysRevMaterials.7.014804} A thin decomposed layer on the surface of the reduced crystals was removed by mechanical polishing.\cite{Planck_PhysRevB.109.235106} X-ray diffraction from the polished surface confirms a phase-pure, highly crystalline LaNiO$_{2}$ composition.
\subsubsection*{NiO}
Commercial powder material from Sigma Aldrich (203882) at 99.99\% trace metals basis. X-ray diffraction confirms a phase-pure compounds (see SI).
\subsubsection*{LiNiO$_{2}$}
Commercial powder material from Sigma Aldrich (757365) with $<$3 $\mu$m particle size (BET) and $\geq$98\% trace metals basis. X-ray diffraction confirms a phase-pure compounds (see SI).
\subsubsection*{Li$_{2}$NiO$_{3}$}
The Ni(OH)$_{2}$ precursor was prepared through a precipitation reaction carried out in a stirred tank reactor (Eppendorf). A 2 M NiSO$_{4}$ solution was pumped into a base solution of 0.4 M NH$_{4}$OH within the reactor. Concurrently, separate solutions of 2 M NaOH (molar ratio NaOH : Ni = 2) and a NH$_{4}$OH (NH$_{4}$OH : TM = 1.2) were pumped into the reactor. A pH of 11 was maintained by the reactor by adjusting the NaOH flow rate. The reaction was stirred for 20 hours at 1000 rpm, where the reactor temperate was maintained at 60 $^{\circ}$C. The Ni(OH)$_{2}$ precipitate was obtained after washing and drying at 80 $^{\circ}$C overnight.

To obtain Li-rich Li$_{2}$NiO$_{3}$ powder, the solid-state preparation method reported by Bianchini et al. was employed. \cite{Reaction_li2nio3} Stoichiometric amounts of Ni(OH)$_{2}$ and LiOH$\cdot$H$_{2}$O were thoroughly mixed through hand grinding and transferred to a tube furnace. To form Li$_{2}$NiO$_{3}$, a pre-heating step at 300 $^{\circ}$C for 12 hours was first applied, followed by further heating at 550 $^{\circ}$C for 24 hours. All heating steps were performed under pure O$_{2}$ flow and heating rates were set to 3 $^{\circ}$C/min. The resulting mixture was then ground and stored in an Ar-filled glovebox before use.

Ogley et al. \cite{Electrochemical_testing_li2nio3} have tested the electrochemical performance, confirmed the material structure to be phase-pure indexed by the C2/m space group, and showed electronic structure of this material to be best described with a stable double ligand hole \textit{d}$^{8}$\underline{L}$^{2}$ state.

%\threesubsection{Second part of experimental section}\\

\subsection*{Single crystalline LiNiO$_{2}$}
\subsubsection*{Synthesis}
Single crystalline (SC) LiNiO$_{2}$ (LNO) was prepared using Ni(OH)$_{2}$, synthesised with the same method as for Li$_{2}$NiO$_{3}$.
% precursor through a controlled precipitation process in a stirred tank reactor. The Ni(OH)$_{2}$ results from a 2 M Ni(SO)$_{4}$ solution, carefully introduced into tank reactor with continuous stirring of 4 M NH$_{4}$OH solution. Concurrently, a 2 M NaOH solution and 0.5 M NH$_{4}$OH (a chelating agent) were added independently. The mixture was vigorously stirred at 1000 rpm for 20 hours, resulting in the formation of pristine Ni(OH)$_{2}$ precipitates. These precipitates were washed thoroughly with deionised water to remove impurities and dried overnight at 80 $^{\circ}$C.

A molten-salt assisted method was employed to obtain single crystal morphology. The Ni(OH)$_{2}$ powder was finely ground in an agate mortar with LiOH.H$_{2}$O and Li$_{2}$SO$_{4}$ in a molar ratio of 1:1.5:0.25. The homogenised mixture was placed in an alumina crucible and subjected to a two-stage heat treatment in a tube furnace under O $_{2}$ atmosphere. The first stage involved heating at 480 $^{\circ}$C for 12 hours, followed by a second stage at 775 $^{\circ}$C for 24 hours. After cooling, the product was washed with deionized water to remove residual lithium species, recovered by centrifugation, and subjected to a final heat treatment at 775 $^{\circ}$C for 6 hours under an O atmosphere to minimize surface degradation. All heating stages were performed with a ramp rate of 5 $^{\circ}$C/min. The material was removed from the furnace at 200 $^{\circ}$C and immediately transferred to a glovebox for storage to preserve its quality.
\subsubsection*{Slurry formulation and electrode preparation}
The slurry formulation and casting of SC-LNO were conducted in a controlled dry room environment (dew point -45 $^{\circ}$C). A 3 grams active material was mixed with commercial carbon black (C65) and polyvinylidene fluoride (PVDF) binder in a weight ratio of 90:5:5. The mixture was homogenised using a thinky mixer at 1300 rpm for 5 minutes. Anhydrous N-methyl-2-pyrrolidone (NMP) (0.9 g) was then added to form a uniform slurry, followed by 15 minutes of mixing to achieve a solid content of 53\%.

The slurry was coated onto a 15 $\mu$m-thick aluminum foil using a 260-$\mu$m doctor blade, ensuring uniform deposition. The coated electrodes were dried under vacuum at 120 $^{\circ}$C overnight. Calendaring was performed using a two-roller compactor at 85 $^{\circ}$C and a roller speed of 1 m/min, resulting in a pressed density of 3.0 g/cm$^{3}$, an areal capacity of 2.57 mAh/cm$^{2}$.

\subsubsection*{Coin cell assembling and electrochemical cycling}\label{sec:LNOCycling}
The assembly of 2325 coin cells (vs. Li$|$Li$^{+}$ as a counter electrode) was conducted in the glovebox. A Celgard 2325 Trilayer microporous membrane of 25 $\mu$m thickness was used as a separator. The cells were filled with 60 $\mu$L of E151 Solvionic electrolyte, comprised of 12.42:30.82:54.76:2 w/w ratio of LiPF$_{6}$, ethylene carbonate (EC), ethyl methyl carbonate (EMC), and vinylene carbonate (VC), respectively. To ensure full wetting of the electrodes, the cells were held at rest at 25 $^{\circ}$C) for 20 hours. Cells underwent one constant-current C/20 (with C = 220 mA/g) charge process to cell potentials (4.0 V, 4.1 V, 4.2 V, 4.3 V, 4.4 V, and 4.6 V) on a Biologic VMP3 potentiostat cycler. The corresponding active mass used to compute specific capacities were 22.15, 23.01, 22.62, 22.5, 22.24, and 22.7 mg.

For each electrochemical condition presented in the manuscript (i.e., each cut-off potential), we assembled and tested three independent coin cells. This approach was adopted to mitigate experimental uncertainties commonly associated with coin cell assembly, particularly electrode misalignment, which can influence overall cell performance. The electrochemical data shown in main text represent the best-performing cell from each set of three. These representative cells were selected based on electrochemical quality (e.g., Coulombic efficiency, voltage profile, and capacity) and alignment with the expected behaviour of LNO at the corresponding state of charge. While variability was observed within some sets, the selected cells follow a consistent trend across the full electrochemical range, reinforcing that they each accurately represent their respective states of charge.

\subsubsection*{Cell disassembling}
After required cycling coin cells were transferred and carefully disassembled within an argon MBraun glovebox (O$_{2} <$ 0.1 ppm, H$_{2}$O $<$ 0.1 ppm) to prevent exposure to atmospheric moisture. This guarantees no air-exposure during cell handling after cycling. Once inside the glovebox, we opened the cells to retrieve the electrodes on which later excess electrolyte was thoroughly washed using dimethyl carbonate (DMC) solvent (sourced from Sigma-Aldrich) with an anhydrous purity of 99 percent. After washing and drying, the cleaned electrodes were used in the SI36917 and SI30201 RPES beamline and the SI33459 Ni L-edge beamline after careful sample loading and transportation that avoids air exposure.

\subsection*{LiMn$_{0.6}$Fe$_{0.4}$PO$_{4}$ electrodes and cells}\label{sc:LMFPcycling}
LiMn$_{0.6}$Fe$_{0.4}$PO$_{4}$ (LMFP64) powder was purchased from Gelon LIB and used to fabricate electrodes following in-house procedures. A mixture of LMFP64 powder, carbon black, and polyvinylidene fluoride (PVDF) pre-dissolved in N-methyl-2-pyrrolidone (NMP) was prepared in a weight ratio of 93.5:3.5:3.5. Additional NMP was incorporated to achieve the desired solid content. The resulting slurry was then applied to carbon-coated aluminium foil, dried overnight at 120 $^{\circ}$C, and calendared to a final density of 2.15 g cm$^{-3}$. To charge the material to a specific state of charge (SoC), half coin cells were assembled with LMFP64 positive electrodes (14.8 mm diameter), a polymer separator (H2325), LP57 electrolyte, and a lithium metal disc as the counter electrode. The cells were then charged galvanostatically to a specified potential using a current of 7.9307 mA/g (equivalent to a C/20 charge rate with and average of 22 mg of active material). Specific capacity reported in the manuscript refer to the active material mass at 21.6, 21.8, 21.5, 22.1, 21.5, and 21.7 mg for the 0\%, 20\%, 40\%, 70\%, and 100\% SoC respectively. 

Similar to the LNO case, we assembled and tested three independent coin cells for each electrochemical condition presented in the manuscript (i.e., each cut-off potential). This approach mitigates experimental uncertainties commonly associated with coin cell assembly. The electrochemical data shown in main text represent the best-performing cell from each set of three. These representative cells were selected based on electrochemical quality (e.g., Coulombic efficiency, voltage profile, and capacity) and alignment with the expected behaviour of LMFP64 at the corresponding state of charge. The selected cells follow a consistent trend across the full electrochemical range, reinforcing that they each accurately represent their respective states of charge.

After charging, the cells were carefully transferred and disassembled in an Ar-filled glovebox (O$_{2} <$ 0.1 ppm, H$_{2}$O $<$ 0.1 ppm), where they were rinsed with DMC and dried. This material was used in the SI35075 RPES beamtime after careful sample loading and transportation that avoids air exposure.

%RPES Ni Ledge SI36917, SI30201
%RPES Fe Ledge SI35075
%Ni L-edge SI33459
\subsection*{Material Characterisation}
\subsubsection*{Resonance Photoemission Spectroscopy}
Resonant Photoemission Spectroscopy (RPES) at Ni \textit{2p} and Fe \textit{2p} absorption thresholds were recorded at beamline I09 DLS across three sessions: Ni RPES in SI30201 and SI36917 beamtimes, and Fe RPES in SI35075 beamtime. For all these sessions, we used a VG Scienta EW4000 detector with a 70 frame/sec CCD camera. The four Ni-based materials in the formed of powder were used in this beamline. LNO half-cells, described in \hyperref[sec:LNOCycling]{the LNO coin cell electrochemical cycling section}, charged to and opened at 4.2 V, 4.4 V, and 4.6 V were used in these beamline. Additionally, LMFP electrodes, described in \hyperref[sc:LMFPcycling]{the LMFP cell section} and charged at 20\%, 40\%, 70\%, and 100\% SoC were also used in this beamline. The total-energy resolution for the measurements at I09 was $<$ 0.2 eV. The calibration of the photon energy was performed by comparing the kinetic energy of the Au \textit{4f} peak. The base pressure during all measurements was less than 2$\times$10$^{-10}$ torr and all measurements were performed at room temperature. After the LNO and LMFP cells were transferred into the argon MBraun glovebox (O$_{2} <$ 0.1 ppm, H$_{2}$O $<$ 0.1 ppm) and excess of electrolyte were washed off with DMC, we prepared the material for the RPES beamtimes inside the WMG glovebox. We mounted the samples on copper plates and placed these plates into the sample cassette provided by the I09 beamline team. This cassette features a sealing system that preserves the argon-inert environment during transport and loading at the I09 facility. It ensures the samples remain protected from environmental contamination and air exposure throughout the process, from the electrochemical cycling to introduction into the spectrometer sampling chamber.

\subsubsection*{XAS L-edge Total Electron Yield and Total Fluorescent Yield}
Ni L-edge X-ray Absorption Spectroscopy were measured in TEY and total fluorescent yield (TFY) mode at the B07 beamline \cite{Grinter:vy5019} at DLS under the SI33459 beamtime. Energy calibration was performed using a NiO reference. The Ni spectra were collected with a energy resolution $<$ 0.15. LNO half-cells, described in \hyperref[sec:LNOCycling]{the LNO coin cell electrochemical cycling section}, charged to and opened at 4.0 V, 4.1 V, 4.2 V, 4.3 V, and 4.4 V were used in these beamline. These electrodes were retrieved, DMC-cleaned, and prepared under Ar-inert conditions in our Ar-filled glovebox (O$_{2} <$ 0.1 ppm, H$_{2}$O $<$ 0.1 ppm) at WMG. We proceeded to load the samples onto sample plates provided by the B07 team. Inside the glovebox, the plates were vacuum-sealed in pouch cells designed to hold the Ar-environment long enough to allow for a secure transportation and loading once at the B07 facilities, guaranteeing no-air exposure during the operation and transport of the material.

\subsubsection*{XAS L-edge SIA Simulations}
\textbf{Ni L-edge simulations} were performed with a parameterised model of a single NiO$_{6}$ octahedron (Oh point group), that contained Ni \textit{2p}, Ni \textit{3d}, and ligand orbitals. The ligand orbitals were defined as linear combinations of oxygen \textit{2p} Wannier-orbitals. The model Hamiltonian consisted of the Coulomb repulsion between (1) two Ni \textit{3d} electrons (including all multiplet effects), (2) a Ni \textit{2p} core and \textit{3d} valence electron (including all multiplet effects), (3) Spin orbit interaction in Ni \textit{3d} and Ni \textit{2p} core level, (4) the onsite energy of the Ni \textit{2p} core orbitals, (5) the orbital dependent onsite energy of the Ni \textit{3d} valence and ligand orbitals, and (6) the hybridisation strength between the Ni \textit{3d} and ligand orbitals. Using this Hamiltonian, XAS excitation was calculated using \href{https://quanty.org/start}{QUANTY} which calculates spectra implemented the Green's function under the dipole approximation. 

Parameter values enter our model in the form of Coulomb interactions, on-site energies, spin orbit interactions, and hopping integrals. The values for these parameters have been quite well established over several decades of core level spectroscopy and other techniques (see \cite{Haverkort_MLFT_2012}, \cite{PhysRevLett.55.418}, and \cite{PhysRevB.33.8060}).  For the monopole Coulomb interaction parameters, we used Udd = 6 eV and Upd = 7 eV. For the ligand-field splitting, we used 10Dq = 0.95 eV between \textit{d} orbitals and 10DqL = 1.44 eV between the Wannier ligand orbitals. For the intra-cluster hopping integrals we used Veg = 3.0 eV and Vt2g = 1.74 eV. Spin orbit interaction parameters were taken as the atomic values for Ni \textit{3d7}, $\xi$2p = 11.3069 eV and $\xi$3d = 0.091 eV. Finally, the multipole Coulomb interaction parameters are taken as 80\% of their atomic Hartree-Fock values for Ni \textit{3d7} [R. Cowan, The Theory of Atomic Structure and Spectra]: F2dd = 10.622, F4dd = 6.636, F2pd = 6.680, G1pd = 5.066, and G3pd = 2.882, all expressed in units of electron volts. A charge transfer energy ($\Delta$) of 4.2 eV for $d^{8}\underline{L}^{0}$ states was selected while using -1.2 eV and -2.6 eV for the $d^{8}\underline{L}^{1}$ and $d^{8}\underline{L}^{2}$, respectively.

\textbf{Fe L-edge simulations} were also performed with a parameterised model of a single FeO$_{6}$ octahedron (Oh point group), that contained Fe \textit{2p} and Fe \textit{3d}. No ligand orbitals are employed as in the ionic picture, O ions only interact elctrostatically with the Fe \textit{3d} orbitals resulting in the expected \textit{t$_{2g}$} and \textit{e$_{g}$} orbitals. The model Hamiltonian consisted of the Coulomb repulsion between (1) two Fe \textit{3d} electrons (including all multiplet effects), (2) a Fe \textit{2p} core and \textit{3d} valence electron (including all multiplet effects), (3) Spin orbit interaction in Fe \textit{3d} and Fe \textit{2p} core level, (4) the orbital dependent onsite energy of the Fe \textit{3d} valence. Using this Hamiltonian, XAS excitation was calculated using \href{https://quanty.org/start}{QUANTY} which calculates spectra implemented the Green's function under the dipole approximation. 

Parameter values enter our model in the form of Coulomb interactions, on-site energies, spin orbit interactions, and hopping integrals. For the ligand-field splitting, we used 10Dq = 1.1 eV between \textit{d} orbitals. Spin orbit interaction parameters were taken as the atomic values for Fe \textit{3d6}, $\xi$2p = 8.2000 eV and $\xi$3d = 0.0520 eV. Finally, the multipole Coulomb interaction parameters are taken as 80\% of their atomic Hartree-Fock values for Fe \textit{3d6} [R. Cowan, The Theory of Atomic Structure and Spectra]: F2dd = 9.8685, F4dd = 6.1335, F2pd = 6.1128, G1pd = 4.5000, and G3pd = 2.5587, all expressed in units of electron volts.

\subsubsection*{DMFT calculations}
To obtain DFT-based Green's functions as the starting point for our DFT+DMFT calculations, we employed the full-potential augmented
plane-wave basis as implemented in the WIEN2K \cite{wien2k}.
For the WIEN2K calculations, we used the largest possible muffin-tin radii, and the basis set plane-wave cutoff was defined by $R_{min}\cdot K_{max} = 9$, where $R_{min}$ is the muffin-tin radius of the O atoms. 

DMFT calculations were performed using the TRIQS/DFTTools modules \cite{aichhorn1, aichhorn2, aichhorn3} based on the TRIQS libraries~\cite{triqs}. We perform DMFT calculations in a basis set of projective Wannier functions as implemented in the dmftproj module of TRIQS. It was also used to calculate the initial occupancy of the correlated orbitals. A projection window of $-10$\,eV to $+26$\,eV was chosen. The large window of unoccupied bands was chosen to account for any hybridisation between Ni $d$ and O $p$ orbitals in the higher energy unoccupied bands, for more accurate charge projections within the $d-dp$ model. All five Ni d orbitals have been treated in the impurity model, whereas the oxygen states have been taken into account as non-interacting.

The Anderson impurity model constructed by mapping the many-body lattice problem to a local problem of an impurity interacting with a bath was solved using the continuous-time quantum Monte Carlo algorithm in the hybridisation expansion (CT-HYB)~\cite{werner06} as implemented in the TRIQS/CTHYB module~\cite{pseth-cpc}. For each DMFT step 150000$\times$128 cycles of warmup steps and 1500000$\times$128 cycles of measures were performed for the quantum Monte Carlo calculations.
We performed one-shot self-consistent DFT+DMFT calculations, using a fully localised limit (FLL) type double-counting correction\cite{helddc}. 
We use a fully rotationally invariant Kanamori Hamiltonian parametrised by Hubbard $U$ and Hund's coupling $J_H$, where we set the intraorbital interaction to $U'=U-2J_H$.
For our DMFT calculations, we used $U$ values ranging from 6 to 9\,eV and $J_H = 0.5$ to $0.75$\,eV to scan the full range of the Metal-Insulator transition. The insulating state was seen to appear at $U = 7$\,eV, $J_H = 0.5$\,eV, hence for $U' = 6$\,eV, which also match with the previous existing values of $U'$ in literature \cite{GenreithSchriever2023, Banerjee2024, lno-dp, nmc, llno}. This value also leads to good agreement of DMFT with experimental results, as we show in this paper.
Real-frequency self-energies have been obtained using the maximum-entropy method of analytic continuation as implemented in the TRIQS/MAXENT module\cite{maxent}. DMFT total and projected DOS have been obtained from the real-frequency self-energies and the post-processing tools of DFTTools.

\subsubsection*{Green's function based simulation of O-K edge XAS}

In our study, we employed the \textsc{feff10} code for the \textit{ab initio} calculation of K-edge XANES. \textsc{feff} employs Green's function-based formulation of the multiple scattering theory to compute the spectra \cite{feff9, feff10}. The X-ray absorption $\mu$ is calculated in a manner similar to Fermi's golden rule when written in terms of the projected photo-electron density of final states or the imaginary part of the one-particle Green's function, $G(r,r';E)$. In terms of the Green's function, $G(r,r';E)$, the absorption coefficient, $\mu$, from a given core level $c$ is given by ref \cite{feff-xas}.
\textcolor{black}{$$\mu = - \frac{1}{\pi}Im \left< c \right| \epsilon_r G(r,r';E) \epsilon_{r'} \left| c \right >$$}
with the Green's function, $G(r,r';E)$ given by
$$G(r,r';E) = \sum_{f} \frac{\Psi_f(r) \Psi_f^*(r')}{E-E_f+i\Gamma}$$
where $\Psi_f$ are the final states, with associated energies $E_f$ and net lifetime $\Gamma$, of a one-particle Hamiltonian that includes an optical potential with appropriate core hole screening. The \textsc{feff} code computes the full propagator $G$ incrementally using matrix factorisation and uses the spherical muffin-tin approximation for the scattering potential.
For self-consistent potential calculations required in the XANES calculation for the Fermi level $E_0$ estimation, a large value of rfms1 \textcolor{black}{(radius of the cluster considered during the full multiple scattering calculation within the self-consistent field loop)} was chosen to be  9 \AA, to have a large number of atoms included in the self-consistent potential calculations. Full multiple scattering (FMS) is required in the XANES calculation, as the multiple scattering (MS) expansion's convergence might not be stable in the XANES calculation. A large rfms \textcolor{black}{(radius of sphere centred on the absorbing atom (real space) or for the unit cell of the crystal (k-space) to compute full multiple scattering calculations.)} value was considered to be  11 \AA,    for proper convergence. The Hedin-Lundqvist self-energy was chosen for the exchange-correlation potential model used for XANES calculation. The random phase approximation (RPA) is used to approximate the core-hole interactions in our K-edge XANES calculations. The default experimental broadening of 0.3eV given by \textsc{feff} was applied.

It is to be noted that the spectral lineshapes obtained from \textsc{feff} are found to be consistent with core-hole DFT spectral calculations using VASP6 \cite{kresse01}; however, \textsc{feff} is more accurate with the calculation of edge energies.

%\section{Conclusion}\label{sec13}
%Not applicable
%Conclusions may be used to restate your hypothesis or research question, restate your major findings, explain the relevance and the added value of your work, highlight any limitations of your study, describe future directions for research and recommendations. 

%In some disciplines use of Discussion or 'Conclusion' is interchangeable. It is not mandatory to use both. Please refer to Journal-level guidance for any specific requirements. 

\section*{Ethics Declarations}
\subsubsection*{Competing Interest}
We declare that none of the authors have competing financial or non-financial interests.
\subsubsection*{Ethics approval and consent to participate}
Not applicable.
\subsubsection*{Consent for publication}
Not applicable.
\subsubsection*{Data availability}
All data plotted in the Article or its Supplementary Information are available via Figshare (https://doi.org/10.6084/m9.figshare.28915994). \cite{DataSet}
\subsubsection*{Materials availability}
Not applicable.
\subsubsection*{Code availability}
Not applicable.
\subsubsection*{Author contribution}
\textbf{G.J.P.F.}: Conceptualization, Project Administration, Supervision, and Writing - Original Draft. \textbf{D.D., M.A., and V.M.}: Data curation, Methodology, Visualization, and Investigation. \textbf{D.D.}: Writing - Original Draft, Formal Analysis, and Validation. \textbf{H.B.}: Writing - Review \& Editing, Formal Analysis, and Validation. \textbf{M.J.W.O., and M.A.}: Writing - Review \& Editing. \textbf{M.H., S.H., P.P., M.I., B.K.}: Resources - LaNiO$_{2}$ single crystal synthesis and material quality control. \textbf{I.M.}: Resources - LiNiO$_{2}$, Li$_{2}$NiO$_{3}$ material synthesis and material quality control. \textbf{S.A.C.:} Resources - LiNiO$_{2}$, Li$_{2}$NiO$_{3}$ material synthesis and material quality control, Funding acquisition. \textbf{P.K.T, T.L., D.C.G., P.F.}: Investigation and Instrumental Resources - guidance in data collection during beamtimes. \textbf{L.F.J.P.}: Funding acquisition, Conceptualization, Project Administration, Supervision, Writing - Review \& Editing


\begin{thebibliography}{10}
\expandafter\ifx\csname url\endcsname\relax
  \def\url#1{\burl{#1}}\fi
\expandafter\ifx\csname urlprefix\endcsname\relax\def\urlprefix{URL }\fi
\providecommand{\bibinfo}[2]{#2}
\providecommand{\eprint}[2][]{\url{#2}}
\providecommand{\doi}[1]{\url{https://doi.org/#1}}
\bibcommenthead

\bibitem{Frith2023}
\bibinfo{author}{Frith, J.~T.}, \bibinfo{author}{Lacey, M.~J.} \&
  \bibinfo{author}{Ulissi, U.}
\newblock \bibinfo{title}{{A non-academic perspective on the future of
  lithium-based batteries}}.
\newblock
  \emph{\bibinfo{journal}{\href{http://dx.doi.org/10.1038/s41467-023-35933-2}{Nature
  Communications}}} \textbf{\bibinfo{volume}{14}} (\bibinfo{year}{2023}).

\bibitem{Paez_2023}
\bibinfo{author}{P\'aez~Fajardo, G.~J.} \emph{et~al.}
\newblock \bibinfo{title}{{Synergistic Degradation Mechanism in Single Crystal
  Ni-Rich NMC//Graphite Cells}}.
\newblock
  \emph{\bibinfo{journal}{\href{https://doi.org/10.1021/acsenergylett.3c01596}{ACS
  Energy Letters}}} \textbf{\bibinfo{volume}{8}}, \bibinfo{pages}{5025--5031}
  (\bibinfo{year}{2023}).

\bibitem{Zhang2022}
\bibinfo{author}{Zhang, M.} \emph{et~al.}
\newblock \bibinfo{title}{{Pushing the limit of 3d transition metal-based
  layered oxides that use both cation and anion redox for energy storage}}.
\newblock
  \emph{\bibinfo{journal}{\href{https://doi.org/10.1038/s41578-022-00416-1}{Nature
  Reviews Materials}}} \textbf{\bibinfo{volume}{7}}, \bibinfo{pages}{522--540}
  (\bibinfo{year}{2022}).

\bibitem{Marie2024}
\bibinfo{author}{Marie, J.-J.} \emph{et~al.}
\newblock \bibinfo{title}{{Trapped O$_{2}$ and the origin of voltage fade in
  layered Li-rich cathodes}}.
\newblock
  \emph{\bibinfo{journal}{\href{https://doi.org/10.1038/s41563-024-01833-z}{Nature
  Materials}}} \textbf{\bibinfo{volume}{23}}, \bibinfo{pages}{818?825}
  (\bibinfo{year}{2024}).

\bibitem{Menon_OxygeReview}
\bibinfo{author}{Menon, A.~S.}, \bibinfo{author}{Ogley, M.~J.},
  \bibinfo{author}{Genreith-Schriever, A.~R.}, \bibinfo{author}{Grey, C.~P.} \&
  \bibinfo{author}{Piper, L.~F.}
\newblock \bibinfo{title}{Oxygen redox in alkali-ion battery cathodes}.
\newblock
  \emph{\bibinfo{journal}{\href{https://doi.org/10.1146/annurev-matsci-080222-035533}{Annual
  Review of Materials Research}}} \textbf{\bibinfo{volume}{54}},
  \bibinfo{pages}{199--221} (\bibinfo{year}{2024}).

\bibitem{Hu2021}
\bibinfo{author}{Hu, E.} \emph{et~al.}
\newblock \bibinfo{title}{{Oxygen-redox reactions in LiCoO$_{2}$ cathode
  without O---O bonding during charge-discharge}}.
\newblock
  \emph{\bibinfo{journal}{\href{https://doi.org/10.1016/j.joule.2021.01.006}{Joule}}}
  \textbf{\bibinfo{volume}{5}}, \bibinfo{pages}{720--736}
  (\bibinfo{year}{2021}).

\bibitem{Zach_LP-topOfChargeC9MH00765B}
\bibinfo{author}{Lebens-Higgins, Z.~W.} \emph{et~al.}
\newblock \bibinfo{title}{{Revisiting the charge compensation mechanisms in
  LiNi$_{0.8}$Co$_{0.2-y}$Al$_{y}$O$_{2}$ systems}}.
\newblock
  \emph{\bibinfo{journal}{\href{http://dx.doi.org/10.1039/C9MH00765B}{Mater.
  Horiz.}}} \textbf{\bibinfo{volume}{6}}, \bibinfo{pages}{2112--2123}
  (\bibinfo{year}{2019}).

\bibitem{Menon_2023PRXEnergy.2.013005}
\bibinfo{author}{Menon, A.} \emph{et~al.}
\newblock \bibinfo{title}{{Oxygen-Redox Activity in Non-Lithium-Excess
  Tungsten-Doped ${\mathrm{Li}\mathrm{Ni}\mathrm{O}}_{2}$ Cathode}}.
\newblock
  \emph{\bibinfo{journal}{\href{https://link.aps.org/doi/10.1103/PRXEnergy.2.013005}{PRX
  Energy}}} \textbf{\bibinfo{volume}{2}}, \bibinfo{pages}{013005}
  (\bibinfo{year}{2023}).

\bibitem{LNO_TrappedD3EE04354A}
\bibinfo{author}{Juelsholt, M.} \emph{et~al.}
\newblock \bibinfo{title}{{Does trapped O$_{2}$ form in the bulk of LiNiO$_{2}$
  during charging?}}
\newblock
  \emph{\bibinfo{journal}{\href{http://dx.doi.org/10.1039/D3EE04354A}{Energy
  Environ. Sci.}}} \textbf{\bibinfo{volume}{17}}, \bibinfo{pages}{2530--2540}
  (\bibinfo{year}{2024}).

\bibitem{Gao2025_Tarsacon_LPsug}
\bibinfo{author}{Gao, X.} \emph{et~al.}
\newblock \bibinfo{title}{{Clarifying the origin of molecular O$_{2}$ in
  cathode oxides}}.
\newblock
  \emph{\bibinfo{journal}{\href{http://dx.doi.org/10.1038/s41563-025-02144-7}{Nature
  Materials}}}  (\bibinfo{year}{2025}).

\bibitem{Qiu2025_Shirley-LPsug}
\bibinfo{author}{Qiu, B.} \emph{et~al.}
\newblock \bibinfo{title}{{Negative thermal expansion and oxygen-redox
  electrochemistry}}.
\newblock
  \emph{\bibinfo{journal}{\href{http://dx.doi.org/10.1038/s41586-025-08765-x}{Nature}}}
   (\bibinfo{year}{2025}).

\bibitem{GENT20201369}
\bibinfo{author}{Gent, W.~E.}, \bibinfo{author}{Abate, I.~I.},
  \bibinfo{author}{Yang, W.}, \bibinfo{author}{Nazar, L.~F.} \&
  \bibinfo{author}{Chueh, W.~C.}
\newblock \bibinfo{title}{{Design Rules for High-Valent Redox in Intercalation
  Electrodes}}.
\newblock
  \emph{\bibinfo{journal}{\href{https://doi.org/10.1016/j.joule.2020.05.004}{Joule}}}
  \textbf{\bibinfo{volume}{4}}, \bibinfo{pages}{1369--1397}
  (\bibinfo{year}{2020}).

\bibitem{GenreithSchriever2023}
\bibinfo{author}{Genreith-Schriever, A.~R.} \emph{et~al.}
\newblock \bibinfo{title}{{Oxygen hole formation controls stability in
  LiNiO$_{2}$ cathodes}}.
\newblock
  \emph{\bibinfo{journal}{\href{http://dx.doi.org/10.1016/j.joule.2023.06.017}{Joule}}}
  \textbf{\bibinfo{volume}{7}}, \bibinfo{pages}{1623?1640}
  (\bibinfo{year}{2023}).

\bibitem{Ogley2025}
\bibinfo{author}{Ogley, M.~J.} \emph{et~al.}
\newblock \bibinfo{title}{{Metal-ligand redox in layered oxide cathodes for
  Li-ion batteries}}.
\newblock
  \emph{\bibinfo{journal}{\href{https://www.sciencedirect.com/science/article/pii/S2542435124004574}{Joule}}}
  \textbf{\bibinfo{volume}{9}}, \bibinfo{pages}{101775} (\bibinfo{year}{2025}).

\bibitem{Walsh2018}
\bibinfo{author}{Walsh, A.}, \bibinfo{author}{Sokol, A.~A.},
  \bibinfo{author}{Buckeridge, J.}, \bibinfo{author}{Scanlon, D.~O.} \&
  \bibinfo{author}{Catlow, C. R.~A.}
\newblock \bibinfo{title}{{Oxidation states and ionicity}}.
\newblock
  \emph{\bibinfo{journal}{\href{http://dx.doi.org/10.1038/s41563-018-0165-7}{Nature
  Materials}}} \textbf{\bibinfo{volume}{17}}, \bibinfo{pages}{958?964}
  (\bibinfo{year}{2018}).

\bibitem{Pavarini:819465}
\bibinfo{editor}{Pavarini, E.}, \bibinfo{editor}{Koch, E.},
  \bibinfo{editor}{van~den Brink, J.} \& \bibinfo{editor}{Sawatzky, G.} (eds)
  \emph{\bibinfo{title}{{{Q}uantum {M}aterials: {E}xperiments and {T}heory}}}
  Vol.~\bibinfo{volume}{6} of \emph{\bibinfo{series}{Schriften des
  Forschungszentrums J\"ulich. Reihe modeling and simulation}}
  (\bibinfo{publisher}{\href{https://juser.fz-juelich.de/record/819465}{Forschungszentrum
  J\"ulich GmbH Zentralbibliothek, Verlag}}, \bibinfo{address}{J\"ulich},
  \bibinfo{year}{2016}).

\bibitem{book_ch3}
\bibinfo{author}{Park, J.-K.}
\newblock \emph{\bibinfo{title}{Cathode Materials}}, Ch.~\bibinfo{chapter}{3},
  \bibinfo{pages}{21--87}
  (\bibinfo{publisher}{\href{https://doi.org/10.1002/9783527650408.ch3}{John
  Wiley \& Sons, Ltd}}, \bibinfo{year}{2012}).

\bibitem{PhysRevB.100.165104}
\bibinfo{author}{Foyevtsova, K.}, \bibinfo{author}{Elfimov, I.},
  \bibinfo{author}{Rottler, J.} \& \bibinfo{author}{Sawatzky, G.~A.}
\newblock \bibinfo{title}{{${\mathrm{LiNiO}}_{2}$ as a high-entropy charge- and
  bond-disproportionated glass}}.
\newblock
  \emph{\bibinfo{journal}{\href{https://link.aps.org/doi/10.1103/PhysRevB.100.165104}{Phys.
  Rev. B}}} \textbf{\bibinfo{volume}{100}}, \bibinfo{pages}{165104}
  (\bibinfo{year}{2019}).

\bibitem{Experiment_Holes_ESRF}
\bibinfo{author}{Jacquet, Q.} \emph{et~al.}
\newblock \bibinfo{title}{{A Fundamental Correlative Spectroscopic Study on
  Li$_{1-x}$NiO$_{2}$ and NaNiO$_{2}$}}.
\newblock
  \emph{\bibinfo{journal}{\href{https://advanced.onlinelibrary.wiley.com/doi/abs/10.1002/aenm.202401413}{Advanced
  Energy Materials}}} \textbf{\bibinfo{volume}{14}}, \bibinfo{pages}{2401413}
  (\bibinfo{year}{2024}).

\bibitem{PhysRevB.109.035139}
\bibinfo{author}{Kothalawala, V.~N.} \emph{et~al.}
\newblock \bibinfo{title}{{Compton scattering study of strong orbital
  delocalization in a ${\mathrm{LiNiO}}_{2}$ cathode}}.
\newblock
  \emph{\bibinfo{journal}{\href{https://link.aps.org/doi/10.1103/PhysRevB.109.035139}{Phys.
  Rev. B}}} \textbf{\bibinfo{volume}{109}}, \bibinfo{pages}{035139}
  (\bibinfo{year}{2024}).

\bibitem{Melot_2013}
\bibinfo{author}{Melot, B.~C.} \& \bibinfo{author}{Tarascon, J.-M.}
\newblock \bibinfo{title}{{Design and Preparation of Materials for Advanced
  Electrochemical Storage}}.
\newblock
  \emph{\bibinfo{journal}{\href{https://doi.org/10.1021/ar300088q}{Accounts of
  Chemical Research}}} \textbf{\bibinfo{volume}{46}},
  \bibinfo{pages}{1226--1238} (\bibinfo{year}{2013}).
\newblock \bibinfo{note}{PMID: 23282038}.

\bibitem{Assat2018}
\bibinfo{author}{Assat, G.} \& \bibinfo{author}{Tarascon, J.-M.}
\newblock \bibinfo{title}{{Fundamental understanding and practical challenges
  of anionic redox activity in Li-ion batteries}}.
\newblock
  \emph{\bibinfo{journal}{\href{http://dx.doi.org/10.1038/s41560-018-0097-0}{Nature
  Energy}}} \textbf{\bibinfo{volume}{3}}, \bibinfo{pages}{373?386}
  (\bibinfo{year}{2018}).

\bibitem{depth_resolvingACSenergy}
\bibinfo{author}{Fantin, R.} \emph{et~al.}
\newblock \bibinfo{title}{{Depth-Resolving the Charge Compensation Mechanism
  from LiNiO$_{2}$ to NiO$_{2}$}}.
\newblock
  \emph{\bibinfo{journal}{\href{https://doi.org/10.1021/acsenergylett.4c00360}{ACS
  Energy Letters}}} \textbf{\bibinfo{volume}{9}}, \bibinfo{pages}{1507--1515}
  (\bibinfo{year}{2024}).

\bibitem{D4EE02398F}
\bibinfo{author}{An, L.} \emph{et~al.}
\newblock \bibinfo{title}{{Distinguishing bulk redox from near-surface
  degradation in lithium nickel oxide cathodes}}.
\newblock
  \emph{\bibinfo{journal}{\href{http://dx.doi.org/10.1039/D4EE02398F}{Energy
  Environ. Sci.}}} \textbf{\bibinfo{volume}{17}}, \bibinfo{pages}{8379--8391}
  (\bibinfo{year}{2024}).

\bibitem{Ans2025}
\bibinfo{author}{Ans, M.} \emph{et~al.}
\newblock \bibinfo{title}{{Operando X-Ray and Postmortem Investigations of
  High-Voltage Electrochemical Degradation in
  Single-Crystal-LiNiO$_{2}$?Graphite Cells}}.
\newblock
  \emph{\bibinfo{journal}{\href{https://advanced.onlinelibrary.wiley.com/doi/abs/10.1002/aenm.202500597}{Advanced
  Energy Materials}}} \textbf{\bibinfo{volume}{15}}, \bibinfo{pages}{2500597}
  (\bibinfo{year}{2025}).

\bibitem{Banerjee2024}
\bibinfo{author}{Banerjee, H.} \& \bibinfo{author}{Morris, A.~J.}
\newblock \bibinfo{title}{{Theoretical approaches to study degradation in
  Li-ion battery cathodes: Crucial role of exchange and correlation}}.
\newblock
  \emph{\bibinfo{journal}{\href{http://dx.doi.org/10.1557/s43578-024-01408-3}{Journal
  of Materials Research}}} \textbf{\bibinfo{volume}{40}},
  \bibinfo{pages}{2?35} (\bibinfo{year}{2024}).

\bibitem{Davis_PhysRevB.25.2912}
\bibinfo{author}{Davis, L.~C.}
\newblock \bibinfo{title}{{Theory of Resonant Photoemission Spectra of 3$d$
  Transition-Metal Oxides and Halides}}.
\newblock
  \emph{\bibinfo{journal}{\href{https://link.aps.org/doi/10.1103/PhysRevB.25.2912}{Phys.
  Rev. B}}} \textbf{\bibinfo{volume}{25}}, \bibinfo{pages}{2912--2915}
  (\bibinfo{year}{1982}).

\bibitem{PhysRevB.53.10372}
\bibinfo{author}{Tjernberg, O.} \emph{et~al.}
\newblock \bibinfo{title}{{Resonant photoelectron spectroscopy on NiO}}.
\newblock
  \emph{\bibinfo{journal}{\href{https://link.aps.org/doi/10.1103/PhysRevB.53.10372}{Phys.
  Rev. B}}} \textbf{\bibinfo{volume}{53}}, \bibinfo{pages}{10372--10376}
  (\bibinfo{year}{1996}).

\bibitem{Haverkort_MLFT_2012}
\bibinfo{author}{Haverkort, M.~W.}, \bibinfo{author}{Zwierzycki, M.} \&
  \bibinfo{author}{Andersen, O.~K.}
\newblock \bibinfo{title}{{Multiplet ligand-field theory using Wannier
  orbitals}}.
\newblock
  \emph{\bibinfo{journal}{\href{https://link.aps.org/doi/10.1103/PhysRevB.85.165113}{Physical
  Review B}}} \textbf{\bibinfo{volume}{85}}, \bibinfo{pages}{165113}
  (\bibinfo{year}{2012}).

\bibitem{Gerard_energyfuels}
\bibinfo{author}{Bree, G.} \emph{et~al.}
\newblock \bibinfo{title}{{Practical Pathways to Higher Energy Density LMFP
  Battery Cathodes}}.
\newblock
  \emph{\bibinfo{journal}{\href{https://doi.org/10.1021/acs.energyfuels.4c06201}{Energy
  \& Fuels}}} \textbf{\bibinfo{volume}{39}}, \bibinfo{pages}{3683--3689}
  (\bibinfo{year}{2025}).

\bibitem{LIU2016109}
\bibinfo{author}{Liu, C.}, \bibinfo{author}{Neale, Z.~G.} \&
  \bibinfo{author}{Cao, G.}
\newblock \bibinfo{title}{{Understanding electrochemical potentials of cathode
  materials in rechargeable batteries}}.
\newblock
  \emph{\bibinfo{journal}{\href{https://www.sciencedirect.com/science/article/pii/S1369702115003181}{Materials
  Today}}} \textbf{\bibinfo{volume}{19}}, \bibinfo{pages}{109--123}
  (\bibinfo{year}{2016}).

\bibitem{Bak2018}
\bibinfo{author}{Bak, S.-M.}, \bibinfo{author}{Shadike, Z.},
  \bibinfo{author}{Lin, R.}, \bibinfo{author}{Yu, X.} \& \bibinfo{author}{Yang,
  X.-Q.}
\newblock \bibinfo{title}{{In situ/operando synchrotron-based X-ray techniques
  for lithium-ion battery research}}.
\newblock
  \emph{\bibinfo{journal}{\href{http://dx.doi.org/10.1038/s41427-018-0056-z}{NPG
  Asia Materials}}} \textbf{\bibinfo{volume}{10}}, \bibinfo{pages}{563?580}
  (\bibinfo{year}{2018}).

\bibitem{Bree2025}
\bibinfo{author}{Bree, G.}, \bibinfo{author}{Majherova, V.},
  \bibinfo{author}{Fiamegkou, E.}, \bibinfo{author}{Moharana, S.} \&
  \bibinfo{author}{Piper, L. F.~J.}
\newblock \bibinfo{title}{{Fast-Charging Lithium-Ion Battery Protocols: LMFP
  Pouch Cells as a Rate Capability Case Study}}.
\newblock
  \emph{\bibinfo{journal}{\href{http://dx.doi.org/10.1149/1945-7111/adb5c7}{Journal
  of The Electrochemical Society}}} \textbf{\bibinfo{volume}{172}},
  \bibinfo{pages}{020526} (\bibinfo{year}{2025}).

\bibitem{Hepting2020}
\bibinfo{author}{Hepting, M.} \emph{et~al.}
\newblock \bibinfo{title}{{Electronic structure of the parent compound of
  superconducting infinite-layer nickelates}}.
\newblock
  \emph{\bibinfo{journal}{\href{http://dx.doi.org/10.1038/s41563-019-0585-z}{Nature
  Materials}}} \textbf{\bibinfo{volume}{19}}, \bibinfo{pages}{381?385}
  (\bibinfo{year}{2020}).

\bibitem{PhysRevB.100.115125}
\bibinfo{author}{Lechermann, F.}, \bibinfo{author}{K\"orner, W.},
  \bibinfo{author}{Urban, D.~F.} \& \bibinfo{author}{Els\"asser, C.}
\newblock \bibinfo{title}{{Interplay of charge-transfer and Mott-Hubbard
  physics approached by an efficient combination of self-interaction correction
  and dynamical mean-field theory}}.
\newblock
  \emph{\bibinfo{journal}{\href{https://link.aps.org/doi/10.1103/PhysRevB.100.115125}{Phys.
  Rev. B}}} \textbf{\bibinfo{volume}{100}}, \bibinfo{pages}{115125}
  (\bibinfo{year}{2019}).

\bibitem{Electrochemical_testing_li2nio3}
\bibinfo{author}{Ogley, M. J.~W.} \emph{et~al.}
\newblock \bibinfo{title}{{Rethinking Oxygen Redox: Does Oxygen Dimerization
  Occur without Oxidation in Li$_{2}$NiO$_{3}$?}}
\newblock
  \emph{\bibinfo{journal}{\href{https://doi.org/10.1021/acsenergylett.4c02031}{ACS
  Energy Letters}}} \textbf{\bibinfo{volume}{9}}, \bibinfo{pages}{4607--4613}
  (\bibinfo{year}{2024}).

\bibitem{PhysRevB.45.1612}
\bibinfo{author}{van Elp, J.}, \bibinfo{author}{Eskes, H.},
  \bibinfo{author}{Kuiper, P.} \& \bibinfo{author}{Sawatzky, G.~A.}
\newblock \bibinfo{title}{{Electronic structure of Li-doped NiO}}.
\newblock
  \emph{\bibinfo{journal}{\href{https://link.aps.org/doi/10.1103/PhysRevB.45.1612}{Phys.
  Rev. B}}} \textbf{\bibinfo{volume}{45}}, \bibinfo{pages}{1612--1622}
  (\bibinfo{year}{1992}).

\bibitem{PhysRevLett.62.221}
\bibinfo{author}{Kuiper, P.}, \bibinfo{author}{Kruizinga, G.},
  \bibinfo{author}{Ghijsen, J.}, \bibinfo{author}{Sawatzky, G.~A.} \&
  \bibinfo{author}{Verweij, H.}
\newblock \bibinfo{title}{{Character of Holes in
  ${\mathrm{Li}}_{x}{\mathrm{Ni}}_{1\ensuremath{-}x}\mathrm{O}$ and Their
  Magnetic Behavior}}.
\newblock
  \emph{\bibinfo{journal}{\href{https://link.aps.org/doi/10.1103/PhysRevLett.62.221}{Phys.
  Rev. Lett.}}} \textbf{\bibinfo{volume}{62}}, \bibinfo{pages}{221--224}
  (\bibinfo{year}{1989}).

\bibitem{PhysRevB.98.075135}
\bibinfo{author}{Dalpian, G.~M.}, \bibinfo{author}{Liu, Q.},
  \bibinfo{author}{Varignon, J.}, \bibinfo{author}{Bibes, M.} \&
  \bibinfo{author}{Zunger, A.}
\newblock \bibinfo{title}{{Bond disproportionation, charge self-regulation, and
  ligand holes in $s\text{\ensuremath{-}}p$ and in $d$-electron ${ABX}_{3}$
  perovskites by density functional theory}}.
\newblock
  \emph{\bibinfo{journal}{\href{https://link.aps.org/doi/10.1103/PhysRevB.98.075135}{Phys.
  Rev. B}}} \textbf{\bibinfo{volume}{98}}, \bibinfo{pages}{075135}
  (\bibinfo{year}{2018}).

\bibitem{PhysRevB.94.195127}
\bibinfo{author}{Green, R.~J.}, \bibinfo{author}{Haverkort, M.~W.} \&
  \bibinfo{author}{Sawatzky, G.~A.}
\newblock \bibinfo{title}{{Bond disproportionation and dynamical charge
  fluctuations in the perovskite rare-earth nickelates}}.
\newblock
  \emph{\bibinfo{journal}{\href{https://link.aps.org/doi/10.1103/PhysRevB.94.195127}{Phys.
  Rev. B}}} \textbf{\bibinfo{volume}{94}}, \bibinfo{pages}{195127}
  (\bibinfo{year}{2016}).

\bibitem{Green_2020_bonddisproportionation}
\bibinfo{author}{Green, R.~J.} \emph{et~al.}
\newblock \bibinfo{title}{{Evidence for bond-disproportionation in LiNiO$_{2}$
  from x-ray absorption spectroscopy}}.
\newblock
  \emph{\bibinfo{journal}{\href{https://arxiv.org/abs/2011.06441}{arXiv}}}
  (\bibinfo{year}{2020}).

\bibitem{PhysRevLett.55.418}
\bibinfo{author}{Zaanen, J.}, \bibinfo{author}{Sawatzky, G.~A.} \&
  \bibinfo{author}{Allen, J.~W.}
\newblock \bibinfo{title}{{Band gaps and electronic structure of
  transition-metal compounds}}.
\newblock
  \emph{\bibinfo{journal}{\href{https://link.aps.org/doi/10.1103/PhysRevLett.55.418}{Phys.
  Rev. Lett.}}} \textbf{\bibinfo{volume}{55}}, \bibinfo{pages}{418--421}
  (\bibinfo{year}{1985}).

\bibitem{SZA_2024_negativeCHtransfer}
\bibinfo{author}{Green, R.~J.} \& \bibinfo{author}{Sawatzky, G.~A.}
\newblock \bibinfo{title}{{Negative Charge Transfer Energy in Correlated
  Compounds}}.
\newblock
  \emph{\bibinfo{journal}{\href{https://doi.org/10.7566/JPSJ.93.121007}{Journal
  of the Physical Society of Japan}}} \textbf{\bibinfo{volume}{93}},
  \bibinfo{pages}{121007} (\bibinfo{year}{2024}).

\bibitem{LCO_ligand_D4CP03759F}
\bibinfo{author}{Asakura, D.} \emph{et~al.}
\newblock \bibinfo{title}{{Elucidation of the Co$^{4+}$ state with strong
  charge-transfer effects in charged LiCoO$_{2}$ by resonant soft X-ray
  emission spectroscopy at the Co L3 edge}}.
\newblock
  \emph{\bibinfo{journal}{\href{http://dx.doi.org/10.1039/D4CP03759F}{Phys.
  Chem. Chem. Phys.}}} \textbf{\bibinfo{volume}{27}},
  \bibinfo{pages}{4092--4098} (\bibinfo{year}{2025}).

\bibitem{Ramachandran2024}
\bibinfo{author}{Ramachandran, H.} \emph{et~al.}
\newblock \bibinfo{title}{{A formal Fe(III/V) redox couple in an intercalation
  electrode}}.
\newblock
  \emph{\bibinfo{journal}{\href{http://dx.doi.org/10.26434/chemrxiv-2024-jhbqx-v2}{ChemRxiv}}}
   (\bibinfo{year}{2025}).

\bibitem{Zuba_ACSEL_li2MnO3}
\bibinfo{author}{Rana, J.} \emph{et~al.}
\newblock \bibinfo{title}{{Quantifying the Capacity Contributions during
  Activation of Li$_{2}$MnO$_{3}$}}.
\newblock
  \emph{\bibinfo{journal}{\href{https://doi.org/10.1021/acsenergylett.9b02799}{ACS
  Energy Letters}}} \textbf{\bibinfo{volume}{5}}, \bibinfo{pages}{634--641}
  (\bibinfo{year}{2020}).

\bibitem{Planck_10.1063/5.0160912}
\bibinfo{author}{Puphal, P.} \emph{et~al.}
\newblock \bibinfo{title}{{Phase formation in hole- and electron-doped
  rare-earth nickelate single crystals}}.
\newblock \emph{\bibinfo{journal}{\href{https://doi.org/10.1063/5.0160912}{APL
  Materials}}} \textbf{\bibinfo{volume}{11}}, \bibinfo{pages}{081107}
  (\bibinfo{year}{2023}).

\bibitem{Planck_PhysRevMaterials.7.014804}
\bibinfo{author}{Puphal, P.} \emph{et~al.}
\newblock \bibinfo{title}{{Synthesis and physical properties of LaNiO$_{2}$
  crystals}}.
\newblock
  \emph{\bibinfo{journal}{\href{https://link.aps.org/doi/10.1103/PhysRevMaterials.7.014804}{Phys.
  Rev. Mater.}}} \textbf{\bibinfo{volume}{7}}, \bibinfo{pages}{014804}
  (\bibinfo{year}{2023}).

\bibitem{Planck_PhysRevB.109.235106}
\bibinfo{author}{Hayashida, S.} \emph{et~al.}
\newblock \bibinfo{title}{{Investigation of spin excitations and charge order
  in bulk crystals of the infinite-layer nickelate LaNiO$_{2}$}}.
\newblock
  \emph{\bibinfo{journal}{\href{https://link.aps.org/doi/10.1103/PhysRevB.109.235106}{Phys.
  Rev. B}}} \textbf{\bibinfo{volume}{109}}, \bibinfo{pages}{235106}
  (\bibinfo{year}{2024}).

\bibitem{Reaction_li2nio3}
\bibinfo{author}{Bianchini, M.} \emph{et~al.}
\newblock \bibinfo{title}{{From LiNiO$_{2}$ to Li$_{2}$NiO$_{3}$: Synthesis,
  Structures and Electrochemical Mechanisms in Li-Rich Nickel Oxides}}.
\newblock
  \emph{\bibinfo{journal}{\href{https://doi.org/10.1021/acs.chemmater.0c02880}{Chemistry
  of Materials}}} \textbf{\bibinfo{volume}{32}}, \bibinfo{pages}{9211--9227}
  (\bibinfo{year}{2020}).

\bibitem{Grinter:vy5019}
\bibinfo{author}{Grinter, D.~C.} \emph{et~al.}
\newblock \bibinfo{title}{{VerSoX B07-B: a high-throughput XPS and ambient
  pressure NEXAFS beamline at Diamond Light Source}}.
\newblock
  \emph{\bibinfo{journal}{\href{https://doi.org/10.1107/S1600577524001346}{Journal
  of Synchrotron Radiation}}} \textbf{\bibinfo{volume}{31}},
  \bibinfo{pages}{578--589} (\bibinfo{year}{2024}).

\bibitem{PhysRevB.33.8060}
\bibinfo{author}{Zaanen, J.}, \bibinfo{author}{Westra, C.} \&
  \bibinfo{author}{Sawatzky, G.~A.}
\newblock \bibinfo{title}{{Determination of the electronic structure of
  transition-metal compounds: 2p x-ray photoemission spectroscopy of the nickel
  dihalides}}.
\newblock
  \emph{\bibinfo{journal}{\href{https://link.aps.org/doi/10.1103/PhysRevB.33.8060}{Phys.
  Rev. B}}} \textbf{\bibinfo{volume}{33}}, \bibinfo{pages}{8060--8073}
  (\bibinfo{year}{1986}).

\bibitem{wien2k}
\bibinfo{author}{Blaha, P.} \emph{et~al.}
\newblock \bibinfo{title}{{WIEN2k: An APW+lo program for calculating the
  properties of solids}}.
\newblock \emph{\bibinfo{journal}{\href{https://doi.org/10.1063/1.5143061}{The
  Journal of Chemical Physics}}} \textbf{\bibinfo{volume}{152}},
  \bibinfo{pages}{074101} (\bibinfo{year}{2020}).

\bibitem{aichhorn1}
\bibinfo{author}{Aichhorn, M.} \emph{et~al.}
\newblock \bibinfo{title}{{Dynamical mean-field theory within an augmented
  plane-wave framework: Assessing electronic correlations in the iron pnictide
  LaFeAsO}}.
\newblock
  \emph{\bibinfo{journal}{\href{https://link.aps.org/doi/10.1103/PhysRevB.80.085101}{Phys.
  Rev. B}}} \textbf{\bibinfo{volume}{80}}, \bibinfo{pages}{085101}
  (\bibinfo{year}{2009}).

\bibitem{aichhorn2}
\bibinfo{author}{Aichhorn, M.}, \bibinfo{author}{Pourovskii, L.} \&
  \bibinfo{author}{Georges, A.}
\newblock \bibinfo{title}{{Importance of electronic correlations for structural
  and magnetic properties of the iron pnictide superconductor LaFeAsO}}.
\newblock
  \emph{\bibinfo{journal}{\href{https://link.aps.org/doi/10.1103/PhysRevB.84.054529}{Phys.
  Rev. B}}} \textbf{\bibinfo{volume}{84}}, \bibinfo{pages}{054529}
  (\bibinfo{year}{2011}).

\bibitem{aichhorn3}
\bibinfo{author}{Aichhorn, M.} \emph{et~al.}
\newblock \bibinfo{title}{{TRIQS/DFTTools: A TRIQS application for ab initio
  calculations of correlated materials}}.
\newblock
  \emph{\bibinfo{journal}{\href{http://www.sciencedirect.com/science/article/pii/S0010465516300728}{Computer
  Physics Communications}}} \textbf{\bibinfo{volume}{204}}, \bibinfo{pages}{200
  -- 208} (\bibinfo{year}{2016}).

\bibitem{triqs}
\bibinfo{author}{Parcollet, O.} \emph{et~al.}
\newblock \bibinfo{title}{{TRIQS: A toolbox for research on interacting quantum
  systems}}.
\newblock
  \emph{\bibinfo{journal}{\href{http://www.sciencedirect.com/science/article/pii/S0010465515001666}{Computer
  Physics Communications}}} \textbf{\bibinfo{volume}{196}}, \bibinfo{pages}{398
  -- 415} (\bibinfo{year}{2015}).

\bibitem{werner06}
\bibinfo{author}{Werner, P.} \& \bibinfo{author}{Millis, A.~J.}
\newblock \bibinfo{title}{{Hybridization expansion impurity solver: General
  formulation and application to Kondo lattice and two-orbital models}}.
\newblock
  \emph{\bibinfo{journal}{\href{https://link.aps.org/doi/10.1103/PhysRevB.74.155107}{Phys.
  Rev. B}}} \textbf{\bibinfo{volume}{74}}, \bibinfo{pages}{155107}
  (\bibinfo{year}{2006}).

\bibitem{pseth-cpc}
\bibinfo{author}{Seth, P.}, \bibinfo{author}{Krivenko, I.},
  \bibinfo{author}{Ferrero, M.} \& \bibinfo{author}{Parcollet, O.}
\newblock \bibinfo{title}{{TRIQS/CTHYB: A continuous-time quantum Monte Carlo
  hybridisation expansion solver for quantum impurity problems}}.
\newblock
  \emph{\bibinfo{journal}{\href{http://www.sciencedirect.com/science/article/pii/S001046551500404X}{Computer
  Physics Communications}}} \textbf{\bibinfo{volume}{200}}, \bibinfo{pages}{274
  -- 284} (\bibinfo{year}{2016}).

\bibitem{helddc}
\bibinfo{author}{Held, K.}
\newblock \bibinfo{title}{{Electronic structure calculations using dynamical
  mean field theory}}.
\newblock
  \emph{\bibinfo{journal}{\href{https://doi.org/10.1080/00018730701619647}{Advances
  in Physics}}} \textbf{\bibinfo{volume}{56}}, \bibinfo{pages}{829--926}
  (\bibinfo{year}{2007}).

\bibitem{lno-dp}
\bibinfo{author}{Banerjee, H.}, \bibinfo{author}{Aichhorn, M.},
  \bibinfo{author}{Grey, C.~P.} \& \bibinfo{author}{Morris, A.~J.}
\newblock \bibinfo{title}{{Insulating behaviour in room temperature
  rhombohedral LiNiO2 cathodes is driven by dynamic correlation}}.
\newblock
  \emph{\bibinfo{journal}{\href{https://dx.doi.org/10.1088/2515-7655/ad7980}{Journal
  of Physics: Energy}}} \textbf{\bibinfo{volume}{6}}, \bibinfo{pages}{045003}
  (\bibinfo{year}{2024}).

\bibitem{nmc}
\bibinfo{author}{Banerjee, H.}, \bibinfo{author}{Grey, C.~P.} \&
  \bibinfo{author}{Morris, A.~J.}
\newblock \bibinfo{title}{{Stability and Redox Mechanisms of Ni-Rich NMC
  Cathodes: Insights from First-Principles Many-Body Calculations}}.
\newblock
  \emph{\bibinfo{journal}{\href{https://doi.org/10.1021/acs.chemmater.4c00928}{Chemistry
  of Materials}}} \textbf{\bibinfo{volume}{36}}, \bibinfo{pages}{6575--6587}
  (\bibinfo{year}{2024}).

\bibitem{llno}
\bibinfo{author}{Banerjee, H.}, \bibinfo{author}{Grey, C.~P.} \&
  \bibinfo{author}{Morris, A.~J.}
\newblock \bibinfo{title}{{Demystifying charge-compensation mechanisms and
  oxygen dimerization in Li-rich Li2NiO3 cathodes}}.
\newblock
  \emph{\bibinfo{journal}{\href{http://dx.doi.org/10.1039/D5TA03794H}{J. Mater.
  Chem. A}}} \bibinfo{pages}{--} (\bibinfo{year}{2025}).

\bibitem{maxent}
\bibinfo{author}{Kraberger, G.~J.}, \bibinfo{author}{Triebl, R.},
  \bibinfo{author}{Zingl, M.} \& \bibinfo{author}{Aichhorn, M.}
\newblock \bibinfo{title}{{Maximum entropy formalism for the analytic
  continuation of matrix-valued Green's functions}}.
\newblock
  \emph{\bibinfo{journal}{\href{https://link.aps.org/doi/10.1103/PhysRevB.96.155128}{Phys.
  Rev. B}}} \textbf{\bibinfo{volume}{96}}, \bibinfo{pages}{155128}
  (\bibinfo{year}{2017}).

\bibitem{feff9}
\bibinfo{author}{Rehr, J.~J.}, \bibinfo{author}{Kas, J.~J.},
  \bibinfo{author}{Vila, F.~D.}, \bibinfo{author}{Prange, M.~P.} \&
  \bibinfo{author}{Jorissen, K.}
\newblock \bibinfo{title}{{Parameter-free calculations of X-ray spectra with
  FEFF9}}.
\newblock
  \emph{\bibinfo{journal}{\href{http://dx.doi.org/10.1039/B926434E}{Phys. Chem.
  Chem. Phys.}}} \textbf{\bibinfo{volume}{12}}, \bibinfo{pages}{5503--5513}
  (\bibinfo{year}{2010}).

\bibitem{feff10}
\bibinfo{author}{Kas, J.~J.}, \bibinfo{author}{Vila, F.~D.},
  \bibinfo{author}{Pemmaraju, C.~D.}, \bibinfo{author}{Tan, T.~S.} \&
  \bibinfo{author}{Rehr, J.~J.}
\newblock \bibinfo{title}{{Advanced calculations of X-ray spectroscopies with
  {\it FEFF10} and Corvus}}.
\newblock
  \emph{\bibinfo{journal}{\href{https://doi.org/10.1107/S1600577521008614}{Journal
  of Synchrotron Radiation}}} \textbf{\bibinfo{volume}{28}},
  \bibinfo{pages}{1801--1810} (\bibinfo{year}{2021}).

\bibitem{feff-xas}
\bibinfo{author}{Rehr, J.~J.} \& \bibinfo{author}{Albers, R.~C.}
\newblock \bibinfo{title}{{Theoretical approaches to x-ray absorption fine
  structure}}.
\newblock
  \emph{\bibinfo{journal}{\href{https://link.aps.org/doi/10.1103/RevModPhys.72.621}{Rev.
  Mod. Phys.}}} \textbf{\bibinfo{volume}{72}}, \bibinfo{pages}{621--654}
  (\bibinfo{year}{2000}).

\bibitem{kresse01}
\bibinfo{author}{Kresse, G.} \& \bibinfo{author}{Furthm\"uller, J.}
\newblock \bibinfo{title}{{Efficient iterative schemes for ab initio
  total-energy calculations using a plane-wave basis set}}.
\newblock
  \emph{\bibinfo{journal}{\href{https://link.aps.org/doi/10.1103/PhysRevB.54.11169}{Phys.
  Rev. B}}} \textbf{\bibinfo{volume}{54}}, \bibinfo{pages}{11169--11186}
  (\bibinfo{year}{1996}).

\bibitem{DataSet}
\bibinfo{author}{P\'aez~Fajardo, G.~J.}
\newblock \bibinfo{title}{{Direct Evidence of Metal-Ligand Redox in Li-ion
  Battery Positive Electrodes}}.
\newblock
  \bibinfo{howpublished}{\href{https://doi.org/10.6084/m9.figshare.28915994}{Figshare-DataSet}}
  (\bibinfo{year}{2025}).

\end{thebibliography}
\end{document}